\documentclass[12pt]{iopart}
\usepackage{graphicx,wrapfig}
\usepackage{iopams}
\usepackage[all,knot,poly]{xy}
\eqnobysec
\newcommand{\C}{\mathbb{C}}
\newcommand{\leftexp}[2]{{\vphantom{#2}}^{#1}{#2}}

\newtheorem{theorem}{Theorem}
\newtheorem{definition}{Definition}

\begin{document}
\title{Topspin networks in loop quantum gravity}
\author{Christopher L Duston}
\address{Physics Department, The Florida State University}
\ead{cduston@fsu.edu}

\begin{abstract}
We discuss the extension of loop quantum gravity to topspin networks, a proposal which allows topological information to be encoded in spin networks. We will show that this requires minimal changes to the phase space, C*-algebra and Hilbert space of cylindrical functions. We will also discuss the area and Hamiltonian operators, and show how they depend on the topology. This extends the idea of ``background independence" in loop quantum gravity to include topology as well as geometry. It is hoped this work will confirm the usefulness of the topspin network formalism and open up several new avenues for research into quantum gravity. 
\end{abstract}

\section{Introduction}\label{sec:intro}
In this paper we explore an idea recently introduced by \cite{DMA}, which incorporates topological information into the existing structures of loop quantum gravity. Essentially, in addition to the geometric data contained in the holonomy, this proposal uses the monodromy to encode topology. In this way we hope to extend loop quantum gravity to the topological realm, so it is clear that ``background independence" applies not just to a geometric background but a topological one as well. As discussed in \cite{Smolin-2005}, general relativity is only background independent (\textit{relational}) with respect to the geometry, and not to the topology, dimension, or differential structure. In the approach discussed in this paper it is possible, at least in the quantum case, to extend gravity to be topologically relational as well. There is also a connection between topological relationalism and the ``problem of time'', which is nicely reviewed in \cite{Anderson}. In addition, this idea may provide a way to take loop quantum gravity to the classical limit, by telling us exactly how to ``fill in the spaces" of a spin network.

The main motivation for this paper is to show that this approach does not modify the basic constructions of loop quantum gravity. A secondary goal is to determine the action of the area and Hamiltonian operators on topspin network states. Thus, will be only interested in details directly relevant for the construction of the phase space, state space, and quantum operators. The basic results are as follows:
\begin{itemize}
\item The structure group $G$  will be replaced with $G\times K$ for finite group $K$.
\item The Lie algebra $\mathfrak{g}$ will be replaced with the universal enveloping algebra $\mathcal{U}(\mathfrak{g})$.
\item The area spectrum is unchanged, but is presented in a form where topological data is now manifest.
\item The action of the Hamiltonian is different due to the equivalent realizations for the topology of the spatial sections.
\end{itemize}

In the rest of this introductory section we will review loop quantum gravity and the topspin network formalism, focusing on the most relevant parts of both. In \S\ref{s:variables} we will discuss the new symmetry which results from the modification of the spin networks and how this changes the phase space of the theory. In \S\ref{s:cyl} we will discuss the state space of the theory, and in \S\ref{s:operators} we present the area operator and explore some differences which result from considering how the Hamiltonian acts on topspin network states. \S\ref{s:conclusion} will conclude and interpret some of our results in the context of topology change in quantum gravity.

We follow the Einstein summation convention, where repeated indices are always summed over. Indices that are not to be summed over will be denoted with a bar, such as in the metric $g(r_{\bar\mu},r_{\bar\mu})$.

\subsection{Loop Quantum Gravity}
We will begin with a brief review of loop quantum gravity (LQG), highlighting specific constructions which we will be discussing later. There are many nice reviews on the subject, both technical \cite{Ash_Review,Han-2007} and non-technical \cite{Ash_EasyReview,Rovelli_Review}. This paper will be borrowing most of the concepts and notations from the texts \cite{Rovelli,TTbook}. Specifically, we will mostly follow the canonical approach rather than the newer covariant approach \cite{Rovelli-covariant-review,Rovelli-covariant-lectures}. This is because the topspin construction in LQG follows from using the spin networks as graphs embedded in $\mathbb{S}^3$, rather than lattice graphs without an explicit embedding which is the natural place to start for the covariant approach. The canonical and covariant approaches result in the same Hilbert space so our choice should only be convenience. 

One begins by rewriting general relativity in the connection formalism on a 1-foliated spacetime $\Sigma \times \mathbb{R}$ where the foliations are parametrized by coordinate time. This results in a constrained Hamiltonian system. Solving these constraints has been the major task of the past twenty years; the first step was the introduction of the \textit{Ashtekar connection} $A$ \cite{Ash_Connection}. The phase space of the theory is given by this connection and the canonical conjugate momentum $E$. These have a formally singular Poisson bracket, and so must be smeared (in one dimension and two dimensions, respectively) with test functions $(c,f)$ on a curve $\gamma$ and a surface $S$. In addition, to make it easier to impose gauge invariance, the holonomy of this $\mathfrak{su}(2)$-connection $h(A,c)$ is used along with the canonically conjugate momentum $E(S,f)$ to define the classical algebra. This technical result provides the background-independent formalism which makes this approach so attractive \cite{TTbook,AshIII}.

The states in LQG are given by \textit{cylindrical functions} $\Psi$ on oriented graphs $\Gamma$ in the spatial slice $\Sigma$. Each edge $e_I$ of a graph is labeled with an irreducible representation $j_{I}$, and each vertex $v_I$ is labeled with an intertwiner
\begin{equation}\iota_I:j_{e_1}\otimes...\otimes j_{e_N}\to j_{e'_1}\otimes...\otimes j_{e'_M},\end{equation}
for $N$ incoming edges and $M$ outgoing edges. The connection on the graph $A$ is the restriction $A|_{\Gamma}$ of the connection on $\Sigma$, and functions of the holonomies $h(A,c)$ can be made gauge invariant by contracting them with the intertwiners. These are the \textit{spin network states} and are given by the collection $(\Gamma,j,\iota)$ (one can also consider gauge-dependent states, but we will not be discussing those). The Hilbert space of cylindrical functions can be defined as a projective limit of the continuous functions on spin networks, with a measure inherited from the Haar measure on $SU(2)$.

This Hilbert space is kinematic only, since we have not solved the Hamiltonian constraint. Geometric operators such as the area and volume can be defined on this Hilbert space, and one of the primary results of LQG is that the area of a surface $S$ intersecting a spin network is quantized. The exact results depends on regularization, but the \textit{restricted spectrum} of the area operator is \cite{TTbook,Rovelli-Smolin}
\begin{equation}A(S)=\frac{l_p^2}{2}\beta\sum_I^N \sqrt{j_I(j_I+1)},\end{equation}
where the surface $S$ intersects the edges $e_I$ of the spin network in $N$ places. $\beta$ is called the \textit{Immirzi parameter}, and can be fixed by considering the semiclassical area of a black hole and comparing to the above result \cite{Domagala-Lewandowski,Meissner}. In the canonical approach the dynamics of the theory are given by the Wheeler-DeWitt equation \cite{DeWitt},
\begin{equation}\label{eq:Wheeler-DeWitt}
\hat{\mathcal{H}}\Psi=0,
\end{equation}
where $\mathcal{H}$ is the Hamiltonian. There are still some ambiguities related to solving the Hamiltonian constraint in LQG \cite{Rovelli,TTbook,QSDII}, but for our purposes we can choose a specific approach and only consider its action on simple spin networks. 

Besides the quantization of area, other results from LQG include the application to early-universe cosmology \cite{Bojowald-book}, black hole physics \cite{Ash-BH}, and recent attempts to couple LQG to fermions and Yang-Mills fields \cite{Han-Rovelli}. We will not be discussing these applications further, but the results of this paper could easily be extended to them.

LQG can be considered a mature theory; much of the framework has been made rigorous and a variety of calculations can be performed. However, there are several features which are still unclear, one of which is the topological nature of the spin networks. Restricting from a smooth three-manifold to an embedded graph naturally trades ``continuous" data in favor of ``discrete" data, and it is unclear how one is to keep track of the topological data from the original manifold. Indeed, since a given graph can be embedded in many topologically different three-manifolds it would appear that topology is completely absent from the quantum theory. This is likely why it has been so difficult to find the classical limit of LQG - the question of how to pass from the discrete data of the graph back to the continuous data of the spatial sections has not been answered.

Topological change in gravity has been studied for quite a while. For smooth manifolds, Morse theory tells us that topological change is associated with critical values of smooth functions \cite{Milnor}. It was also shown by \cite{Geroch} that topology changing manifolds with Lorentzian metrics must have closed timelike curves. A similar ``no go'' situation occurs in linearized gravity, where topology change is associated with infinite particle and energy production \cite{Anderson-DeWitt}. These results suggest that classically, topology change may be forbidden. On the other hand, we should expect that in a background-independent formulation of quantum gravity, one should not \textit{impose} a topology, in the same way that one should not impose a background field \cite{Smolin-2005}. Although background independence is usually formulated in terms of geometric independence (\textit{i.e.} the metric), presupposing a topology should be just as offensive. 

It is sometimes remarked that LQG does not actually describe the original spatial section (usually taken to be a topological three-sphere), but rather a space with a large number of holes given by the topological structure of the spin network \cite{TTbook}. Since the Hamiltonian operator $\hat{\mathcal{H}}$ can add edges to a graph $\Gamma$, it can change the hole structure of the graph, and thus of $\Sigma$. However, this does not solve the problem of being able to pass into a classical limit to check the results of the theory, since (as discussed above), we can easily embed a graph with a large first Betti number into a three-manifold with a much smaller, or possibly trivial, first homology group. 

An alternative idea comes from the Borde-Sorkin conjecture, which avoids the creation of infinite energy due to topology change by restricting to only causally continuous spacetimes \cite{Dowker2002,Dowker2005}. This is known as causal set theory, and it allows for topology change by a full discretization of spacetime, which is then built up to be a model for a smooth manifold. This approach brings along many difficulties but one surprising result; it seems to predict the order of magnitude of the cosmological constant \cite{Sorkin}. This approach can be thought of as ``as discrete as possible", removing all other structures except for a partial ordering between spacetime points.

This paper proposes to incorporate topology change into quantum gravity by adding topological data to the already existing spin networks. In this way we can benefit from the extensive theoretical work that has already been done in LQG without further discretizing spacetime as in the causal set approach. We discuss this in detail in the next subsection.

\subsection{Topspin Networks}
First we recall a few definitions \cite{PS}:

\begin{definition} A continuous map $p:M\to B$ between $q$-dimensional manifolds $M$ and $B$ is said to be a \textbf{covering map} if for every point $b\in B$ there exists a neighborhood $U(b)$ such that restriction of $p$ to each connected component of $p^{-1}(U(b))$ is a homeomorphism to $U(b)$. $M$ is the \textbf{covering space} and $B$ is the \textbf{base space}. If the fiber of the covering map has $n$ points we call $p$ an \textbf{n-fold cover}. A \textbf{branched covering map} is a covering map except at a finite number of points $\{b_1,...,b_m\}\in B$ such that $p^{-1}(\{b_1,...,b_m\})$ is a finite set. The set of points $\{b_1,...,b_m\}$ is called the \textbf{branch locus}, and its complement $M-p^{-1}(\{b_1,...,b_m\})$ is again a covering space. We will call $M$ a \textbf{branched covering space}.
\end{definition}  

Intuitively, a covering space of $B$ is just a discrete set of homeomorphic copies (\textit{sheets}) of $B$ which map back to $B$ under $p$. A branched covering space is a covering space where the sheets can collide with each other over the branch locus.

Topspin networks were first introduced by \cite{DMA}. That paper was motivated by an interesting mathematical fact, first discovered by Alexander:

\begin{theorem} \cite{Alexander, Knots} Every orientable closed manifold of dimension $q$ is a branched covering of $\mathbb{S}^q$, branched along a $(q-2)$ subcomplex.
\end{theorem}

In other words, one can construct any closed $q$-dimensional manifold by gluing together pieces of $q$-spheres along $(q-2)$-dimensional submanifolds. We will mostly be interested in the $q=3$ case, where the branching occurs along graphs, but the $q=4$ is very relevant for spinfoams \cite{DMA}. For three-manifolds, this theorem can be strengthened to three-fold covers along a knot (the Hilden-Montesinos theorem) \cite{Hilden, Montesinos}, although here we consider the general case for a covering of any order. 

Specifying now to $q=3$, a branched covering of order $n$ along an embedded graph $\Gamma\hookrightarrow\mathbb{S}^3$ is completely determined by a representation of the fundamental group $\sigma:\pi_1 (\mathbb{S}^3 \setminus \Gamma)\to S_n$. This gives the classes of loops in the graph complement as elements of the permutation group. Ambient isotopy classes of these graphs can be represented in two dimensions with a \textit{planar diagram}. This is a choice $D(\Gamma)$ for a representative $\Gamma$ of the class where the neighborhood of each node is a set of arcs coming from the node (see \cite{Kauffman} for more background on embedded graphs in three-dimensional space).

We can use planar diagrams to give a presentation of the branched covering spaces from the above theorem: given a planar diagram $D(\Gamma)$, the representation $\sigma$ is determined by assigning elements $\sigma_I\in S_n$ along the arcs of $D(\Gamma)$. Each $e_I$ tells us how to exchange covers over the branch locus. These elements must obey the \textit{Wirtinger relations}. At crossings with an overcrossing edge $K$ and undercrossing arcs $I$ and $J$,
\begin{eqnarray*}
\sigma_J&=\sigma_K \sigma_I\sigma_K^{-1},\\
\sigma_J&=\sigma_K^{-1}\sigma_I\sigma_K.
\end{eqnarray*}
The first relation is for negatively-oriented crossings, the second is for positively-oriented crossings. There are also relations to be satisfied at the vertices,
\begin{equation}\prod _I \sigma_I \prod _J \sigma_J^{-1}=1,\end{equation}
for incoming arcs $I$ and outgoing arcs $J$. These relations encode how the sheets of the covering are stitched together so that $\pi_1(\mathbb{S}^3 \setminus \Gamma)$ is appropriately represented. 

The idea of a topspin network is to identify the branched covers, decorated with permutation labels $\sigma_I$ under the Wirtinger relations, with the usual spin networks of LQG, ensuring that the spin and topological labels are compatible. In this way the spin networks will not only encode all the geometric information of the gravitational field but the topological information as well.

\begin{figure}\begin{center}
\includegraphics[scale=0.5]{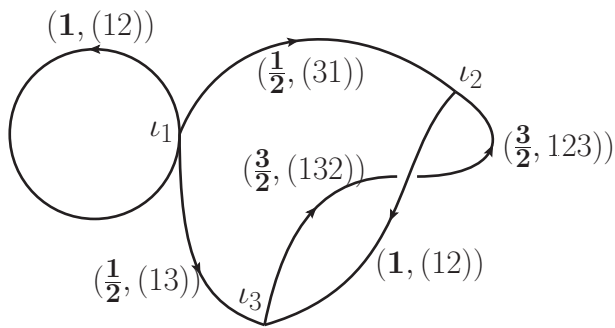}
\caption{A sample topspin network of a three-fold branched cover, with geometric (spin) labels $\vec{j}$, intertwiners $\vec{i}$ and topological labels $\vec{\sigma}$.}\label{fig:Topspin_Ex}
\end{center}\end{figure}

\begin{definition} A \textbf{topspin network} over a compact Lie group $G$ is a tuple $(\Gamma,j,\iota,\sigma)$ of data consisting of
\begin{itemize}
\item a spin network $(\Gamma,j,\iota)$ with an embedded graph $\Gamma\hookrightarrow\mathbb{S}^3$,
\item a representation $\sigma:\pi_1(\mathbb{S}^3\setminus \Gamma)\to S_n$ given by an assignment of $\sigma_I\in S_n$ to each arc $e_I$ in $D(\Gamma)$.
\end{itemize}
\end{definition}
See figure \ref{fig:Topspin_Ex} for an example topspin network. This definition is identical to the original one of \cite{DMA}, although we will be considering only irreducible representations $\rho$ since we want to directly compare this theory to LQG. Under covering moves (which we discuss shortly) one can actually get reducible representations starting with irreducible ones. However, as discussed in \cite{DMA}, there is a trivial equivalence one gets by demanding that the compositions of adjacent intertwiners are the same, which can be removed by considering the representations as irreducible.

\begin{figure}
\begin{center}
\includegraphics[scale=0.4]{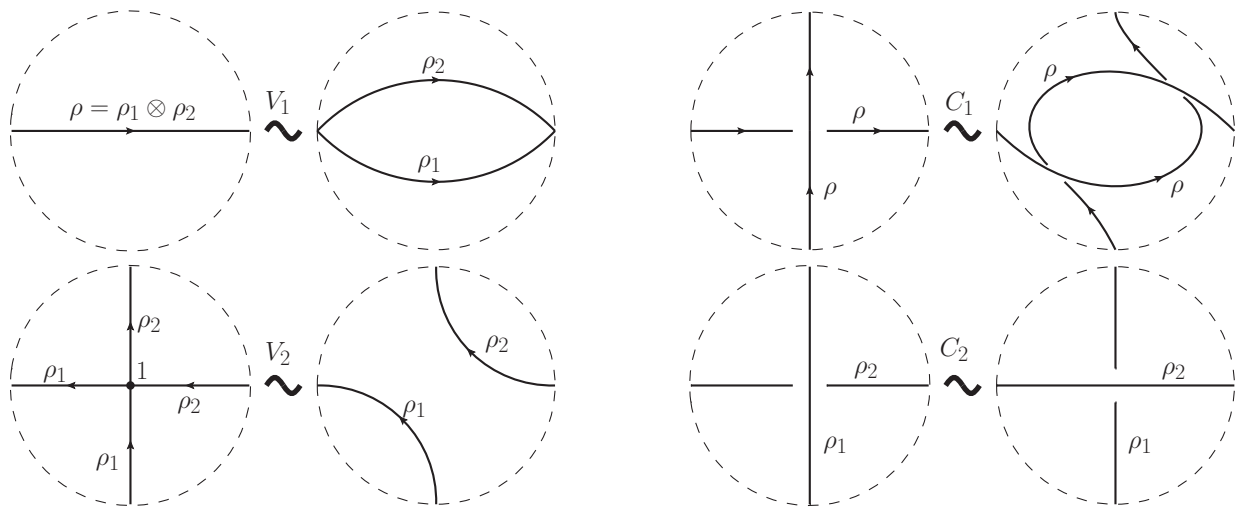}
\caption{Geometric covering moves.}
\label{fig:covering_moves}
\end{center}
\end{figure}

Along with the addition of topological labels, we will have a set of \textit{covering moves} which can modify the labeled branch locus $(\Gamma,\sigma)\to(\Gamma',\sigma')$ but leave the branched covering space unchanged \cite{Covering}. In addition, we will need the geometric labels $(\rho,\iota)$ to be compatible under the change as well. These \textit{geometric covering moves} were found in \cite{DMA}, and we reproduce them here in figure \ref{fig:covering_moves}. These are basically extensions of the Reidemeister moves to our decorated graphs, so that under the geometric covering moves the geometry and topology of the spatial sections do not change. It is also easy to see that we can choose that the number of sheets of the cover does not change under these moves (\textit{stabilization}), since we can always add another sheet branched over a trivial link with a simple permutation of two sheets. 

Since Alexander's theorem can be extended to dimension four as well, this basic construction can also be extended to spin foams \cite{DMA}. One can then define \textit{topspin foams}, which are embedded 2-complexes in $\mathbb{S}^3\times [0,1]$ with spin and topological labeling. There are similar node relations and covering moves, and any four-manifold can be described with such an embedded surface \cite{Piergallini}. We will not discuss the extension of the construction in this paper to topspin foams, as our goal here is to just show that LQG can be consistently defined at least on topspin networks.  

\section{Loop Quantum Gravity as $SU(2)\times S_n$ Gauge Theory}\label{s:variables}
\subsection{Phase Space}\label{s:ADM}
To construct the phase space of the theory, we begin by describing the spatial sections as branched coverings of the three-sphere $p:\Sigma\to \mathbb{S}^3$, where the branching is over a graph $\Gamma\subset \mathbb{S}^3$ and we have a representation $\sigma:\pi_1(\mathbb{S}^3\setminus\Gamma)\to S_n$. For example, a three-fold cover over an oriented graph $\Gamma$ with an element $(1)(23)\in S_3$ assigned to an edge $e\in\Gamma$ means that as one crosses the preimage $p^{-1}(e)$, the second and third cover exchange while the first does not. The action of $S_3$ on the covers (the \textit{deck transformation}) induces a symmetry on the charts of the spatial section which was not present before. 

In a covering space, the preimage $p^{-1}(U)$ of a chart $U\subset \mathbb{S}^3$ is a direct product of a finite number of homeomorphic copies of $U$. In a branched covering space, some of these charts are identified with each other over the branch locus. Intuitively, what we are doing is noticing that there are some charts in the spatial section which are homeomorphic to each other, and folding the section so these charts are all in the inverse image of a chart in the base space. This can always be done by Alexander's theorem, and we assign permutation labels that satisfy the Wirtinger relations to ensure the topology of the spatial section is unchanged. Now, since this is generally a \textit{branched} covering, there are curves in the spatial section which can cross between sheets of the covering. At the branch locus of these crossings, the inverse images of the open sets need to be identified. This symmetry appears to be particularly relevant for LQG since the identification occurs exactly at the branch loci, which is where all the dynamics of the quantum theory actually take place.

More precisely, there are canonical coordinates $(x^a,\mu):=x_\mu^a$ over $\mathbb{S}^3 \setminus \Gamma$ where $x^a\in\mathbb{S}^3\setminus \Gamma$
and $\mu$ labels the sheets of the cover. We can extend these to the branch loci by replacing $\mu$ with the equivalence class $[\mu]$ of sheets that are identified with the $\mu$th one under the action of $S_n$ and extending the coordinates $x^a$ to $x^a\in\mathbb{S}^3$. In other words, in a neighborhood above an edge $e_I$ we have
\begin{equation}\label{eq:deck}
x^a_{\mu}\sim x^a_{\nu}\mbox{ if }x^a_{\mu}=\sigma(e_I)x^a_{\nu},~\sigma(e_I)\in S_n.
\end{equation}
Here one can write in cycle notation $\sigma(e_I)=(\mu \nu)$. This is generic since elements of $S_n$ can always be written as products of transpositions for $n>2$. Where transpositions are explicitly indicated, a result on any element $\sigma(e_I)\in S_n$ can be easily found by iteration. 

This symmetry in the coordinates corresponds to a symmetry of the spatial metric $q_{ab}$ by restriction to each edge. Denote points $z^a$ as those which are the restrictions of the coordinates to the branch locus, $z^a=x^a|_\Gamma$. These extend to the cover as above $(z^a,\mu):=z^a_\mu$, where $[\mu]$ is the equivalence class of sheets which are identified with the $\mu$th one. Now within in each equivalence class, the metric should be invariant since the physical position in space is the same:
\begin{equation}\label{eq:restricted_deck}
\fl q(r_{\bar\nu},r_{\bar\nu})=q((\bar\nu \bar\mu)r_{\bar\mu},(\bar\nu \bar\mu)r_{\bar\mu})=q(r_{\bar\mu},r_{\bar\mu}),~(\bar\nu \bar\mu)\in S_n,~r_{\bar\mu},r_{\bar\nu}\in T\Sigma,~\nu\sim\mu.
\end{equation}
Then the components of the metric transform like
\begin{eqnarray*}
q_{ab}z^a_{\bar\nu}z^a_{\bar\nu}&=q_{ab}(\bar\nu \bar\mu)z^a_{\bar\mu}(\bar\nu \bar\mu)z^b_{\bar\mu}\\
&=(\bar\nu \bar\mu)(\bar\mu \bar\nu)q_{ab}z^a_{\bar\mu} z^b_{\bar\mu}=q_{ab}z^a_{\bar\mu}z^a_{\bar\mu}.
\end{eqnarray*}
To do this we have induced a right action from the given left action via
\begin{equation}(\bar\mu \bar\nu)z^a_{\bar\nu}=z^a_{\bar\nu}(\bar\mu \bar\nu)^{-1}=z^a_{\bar\mu}.\end{equation}
This is natural and well-defined since all permutations have inverses. While this transformation must be valid over the branch locus, it does not need to hold for general neighborhoods of $\Gamma$, as in (\ref{eq:deck}). 

This point deserves some attention. If we were require that the metric is not just the same on the branch locus but across neighborhoods of the branch locus as in (\ref{eq:deck}), we are identifying not just the geometry of one-dimensional submanifolds of our spatial sections but the geometry of three-dimensional submanifolds. It is not clear what the physical interpretation of this might be, but there is nothing to suggest that it is forbidden for the spatial sections of our universe to have ``distinct geometry everywhere''. In fact, taking the stance that our universe \textit{cannot} have the same geometry anywhere sounds more outrageous than there are locations where it \textit{must} be the same. One can even think of examples, such as a spherically symmetric spacetime with a gravitational wave propagating outward from $r=0$. Then at every constant $t$ slice the geometric perturbation (the metric) would be the same on the wavefront.

However, this simple argument fails when one considers the metric as a dynamical field. The full deck transformation symmetry suggests that dynamical information could propagate instantaneously between the sheets, since a perturbation on one must be transmitted to the other via (\ref{eq:deck}). When we take the more limited deck transformation symmetry of (\ref{eq:restricted_deck}), we avoid this problem  - we only need the geometry to be the same on a set of measure zero in the spatial sections. This should have no effect on the classical dynamics of the theory since they will not contribute directly to the action. Of course, since the dynamics of LQG occur \textit{only} on the graphs, they will contribute directly to the quantum theory.

In LQG we are interested in the tetrad field $e^i_a$, which is related to the metric by $q_{ab}:=\delta_{jk}e^j_ae^k_b$ \cite{TTbook}. This relation is invariant under $SO(3)$ transformations, and one can canonically associate these fields to $\mathfrak{su}(2)$-valued 1-forms. Given how the metric transforms we can determine how these fields transform under $S_n$:
\begin{equation}q(r_{\bar\mu},r_{\bar\mu})=\delta_{jk}e^j((\bar\mu \bar\nu)r_{\bar\nu})e^k((\bar\mu \bar\nu)r_{\bar\nu}).\end{equation}
This suggests that $S_n$ acting on the coordinates induces an action on the fields,
\begin{equation}e^j(r_\nu)=\sigma(e_I)e^j(r_\mu),~ \sigma(e_I)=(\nu \mu)\in S_n.\end{equation}
Thus our fields must transform under $SU(2)\times S_n$, and should be 1-forms which take values in an algebra associated to this group structure. We do not want to simply use the Lie algebra of this group, since the Lie algebra of a discrete group is 0 and we will lose all the information regarding the sheets of the cover. In addition, Lie algebras are not associative algebras (Lie algebras have commutators but no product structure), so we can not take a tensor product of $\mathfrak{su}(2)$ with some associative algebra naturally associated to the permutation group. 

But given a Lie algebra $\mathfrak{g}$, one can always find the universal enveloping algebra $\mathcal{U}(\mathfrak{g})$, which is a unital associative algebra. This algebra is the quotient of the free algebra of $\mathfrak{g}$ by elements of the form $[x,y]=x\cdot y - y\cdot x~\forall x,y\in \mathfrak{g}$. For the algebraic structure associated to $S_n$ we will take the group algebra $\mathbb{C}S_n$. Elements $a\in \mathbb{C}S_n$ can be written like $a=f(\sigma_\mu)\sigma^\mu$, where $f:S_n\to \mathbb{C}$ and $\sigma^\mu\in S_n$, indexed by $1\leq \mu \leq n!$. Multiplication in this algebra is given by the convolution product
\begin{equation}(f_1 *f_2)(\sigma)=\sum_{\sigma=\sigma_1\sigma_2}f_1(\sigma_1)f_2(\sigma_2).\end{equation} 
A convenient $n$ dimensional representation for the symmetric group is the permutation representation $\rho:S_n\to GL(\mathbb{C}^n)$ which permutes the basis elements of $\mathbb{C}^n$. We will denote $\rho(\sigma_\mu)=\rho_\mu$ for basis elements in this representation. There is also a natural inner product on this algebra given by
\begin{equation}\label{eq:CS_product}
(a,b):=\frac{n^2}{|S_n|}\sum_{\sigma,\sigma'\in S_n}a^*(\sigma)b(\sigma')\delta_{\sigma\sigma'},~\forall a,b\in \C S_n.
\end{equation}
The choice of scaling $n^2/|S_n|$ is at this point arbitrary; we will explain our choice when we discuss the area operator in \S\ref{s:operators}.

The algebraic structure for the fields will  be the tensor product of these two algebras:
\begin{equation}\mathcal{A}:=\mathcal{U}(\mathfrak{su}(2))\otimes \mathbb{C}S_n,\qquad e_a\in \Gamma (\mathbb{S}^3,\mathcal{A}).\end{equation}
In other words, our fields $e$ are now $\mathcal{A}$-valued one-forms. The Poincar\'e-Birkoff-Witt theorem \cite{Dixmier} gives us a basis for $\mathcal{U}(\mathfrak{su}(2))$, which are polynomials in the basis $\{T_i~|~1\leq i \leq 3\}$ of $\mathfrak{su}(2)$,
\begin{equation}\{T_1^aT_2^bT_3^c~|~a,b,c \in \mathbb{N}\}.\end{equation}
This is an infinite dimensional vector space with basis elements labeled $\tau_{abc}$. If we order the basis lexicographically we can label it with a single index $\tau_i$ and write elements of $\mathcal{A}$ as $a^{i\mu} \tau_i\otimes\rho_\mu$. The spatial metric is then
\begin{equation}q_{ab}=\delta_{ik}\delta_{\mu\nu} e^{i\mu}_a e^{k\nu}_b.\end{equation}
The conjugate variables to our coordinate fields $e^{i\mu}_a$ will be the fields $K^{i\mu}_a$ associated with the extrinsic curvature
\begin{equation}-sK_{ab}:=\delta_{ij}\delta_{\mu\nu}K^{i\mu}_{(a}e^{j\nu}_{b)}.\end{equation}
It is customary to define the scaled coordinate fields as $E^a_{j\mu}:=\sqrt{det(q)}e^a_{j\mu}$.
Here we are raising and lowering indices with delta functions and have defined the fields so that $e^a_{j\mu}e_a^{i\nu}=\delta^i_j\delta_\mu^\nu,~e^a_{j\mu}e^{j\nu}_b=\delta^a_b\delta^\nu_\mu$, and $e^a_{j\mu}e_b^{i\mu}=\delta^a_b\delta^i_j$. We can equip these fields with a Poisson bracket by using the commutator on the basis elements:				
\begin{equation}\label{eq:bracket}
\{A_a,B_b\}=\{A_a^{j\mu}\tau_j\otimes \rho_\mu,B_b^{i\nu}\tau_i \otimes \rho_\nu\}=\{A_a^{j\mu},B_b^{i\nu}\}[\tau_j\otimes \rho_\mu,\tau_i\otimes\rho_\nu],
\end{equation}
where the Poisson bracket is defined with functional derivatives
\begin{equation}\label{eq:Poisson}
\{G,G'\}=\kappa\int_{\sigma}\rmd ^3 x'\Biggl[ \frac{\delta G}{\delta K_a^{j\nu}(x')}\frac{\delta G'}{\delta E^a_{j\nu}(x')}-\frac{\delta G'}{\delta K_a^{j\nu}(x')}\frac{\delta G}{\delta E^a_{j\nu}(x')}\Biggr].
\end{equation}
With this we have our phase space $(K^{j\mu}_a,E^{b}_{i\nu})$ equipped with the standard symplectic structure,
\begin{eqnarray}\label{eq:poisson}
\{E^{a}_{j\mu}(x),E^b_{k\nu}(y)\}=\{K^{j\mu}_a(x),K^{k\nu}_b(y)\}=0,\nonumber\\
\{E^a_{i\mu},K^{j\nu}_{b}\}=\frac{\kappa}{2}\delta^a_b\delta^j_i\delta^\mu_\nu\delta(x,y),
\end{eqnarray}
where $\kappa$ is the coupling constant. This is the same symplectic structure as the usual LQG theory except for the extra index, so the ADM variables would be reproduced in the same manner (see \cite{TTbook} for this calculation).

The enlargement of the algebra to $\mathcal{U}(\mathfrak{su}(2))$ is a direct consequence of using discrete symmetry groups in a physical theory. We can no longer describe the symmetry of the system using Lie algebras because they do not ``see'' discrete groups (we elaborate on this in the next section). Thus, we are forced to work with associative algebras, and the universal enveloping algebra is a natural way to connect Lie algebras and associative algebras. At any stage in our construction one could restrict to just the first-order elements $\mathfrak{su}(2)\subset\mathcal{U}(\mathfrak{su}(2))$ to recover the original Lie algebra, but in doing this one loses all the algebraic structure of the theory - including the bracket (\ref{eq:bracket}). So for full consistency and generality we do not want to do this.

In fact, this enlargement is completely in line with what is usually done in quantum theory; that is, promoting the classical fields to operators. This is done in LQG by writing the area operator as a product of the $\mathfrak{su}(2)$ fields, which we will discuss in more detail in \S\ref{s:operators}. Since differential operators naturally have an algebraic structure isomorphic to a universal enveloping algebra, we are simply promoting our fields to operators immediately and giving them the appropriate algebraic structure.

\subsection{The Canonical Transformation and Classical Algebra}\label{s:2.2}
We will now give a canonical transformation of the Poisson algebra (\ref{eq:poisson}), following the usual path of canonical quantization of the gravitational field. The spin connection on $\mathcal{U}(\mathfrak{su}(2))$ is the usual one for $\mathfrak{su}(2)$ extended to the infinite polynomial basis $\tau_j$,
\begin{equation}
\Gamma_{a}^{jk}=-e^{bk}(\partial_ae^j_b-\Gamma^c_{ab}e^j_c),~j,k\in \mathbb{N}.
\end{equation}
In the usual manner we can rewrite this as $\Gamma_a^{jk}=\Gamma^l_a[M_l]^{jk}$, where the $M_l$ are the matrices of the defining representation of $\mathcal{U}(\mathfrak{su}(2))$; in the usual LQG case these would be $\epsilon_{ijk}$, representing the adjoint of $\mathfrak{su}(2)$. This expression appears somewhat formal since $\mathcal{U}(\mathfrak{su}(2))$ is an infinite-dimensional vector space, but by virtue of the Poincar\'e-Birkoff-Witt theorem all the usual index manipulations are perfectly well-defined.

By way of analogy, our canonical transformation will be 
\begin{equation}\label{eq:connection}
A_a^{k\nu}:=\Gamma_a^{k}\rho^{\nu}+\beta K_a^{k\nu},~\tilde{E}^b_{k\nu}=\beta^{-1}E^b_{k\nu}.
\end{equation}
Here $\beta$ is the Immirzi parameter, which we take to be real. The motivation for the above comes from the original Ashtekar connection, combined with the fact that we want our connection to be defined in the same space $\mathcal{A}$ as our fields. It is easy to see that since the spin connection is an order-zero homogeneous polynomial in $E^b_{k\nu}$, this transformation is canonical and $(\tilde{E},A)$ is a Poisson algebra.

We now need a choice of classical algebra to use for the quantum theory, which will be inspired by the usual LQG case in which one chooses the holonomy and the canonical fields $\tilde{E}$. This choice is motivated by the fact that the holonomy around a closed loop is a gauge invariant, while the connection is not. For a curve $c:[0,1]\to \Sigma$, the holonomy $h(A,c)$ is defined via
\begin{equation}\label{eq:holo}
\frac{\rmd}{\rmd s}h(A,c(s))=h(A,c(s))A(c(s)).
\end{equation}
This has the formal solution $h(A,c(s))=\mathcal{P}\exp (\int_c A(c(s)))$, but it is not clear what the exponential map should be in our case since $\mathcal{A}$ is no longer a Lie algebra. Observe that for a general Lie group $G$, the exponential map has domain $T_1G$, the tangent space at the identity. The tangent space for our Lie group is isomorphic to $T_1 SU(2)$ since $S_n$ is a finite group and has tangent space 0. This suggests that our exponential map should actually be the identity on $\mathbb{C}S_n$ and the usual exponential on $\mathcal{U}(\mathfrak{su}(2))$. This is reasonable because the holonomies (and derivatives) should exist in the continuous directions, not in the finite ones since the tangent space there is trivial. 

To  implement this idea, we can consider the algebra $\mathcal{U}(\mathfrak{g})\otimes \C K$ for Lie algebra $\mathfrak{g}=Lie(G)$ and finite group $K$ to be functions $f\in C(K,\mathcal{U}(\mathfrak{g}))$ with product
\begin{equation}(f_1\star f_2)(k)=\sum_{k=k_1k_2}f_{1}^i(k_1)f_{2}^j(k_2)\tau_i\tau_j.\end{equation}
Here we are writing $f(k)=f^i(k)\tau_i,~\forall k\in K$ and $\tau_i$ is a basis element for $\mathcal{U}(\mathfrak{g})$. Now for the connection $A\in C(K,\mathcal{U}(\mathfrak{g}))$, write the holonomy equation for an element $k\in K$ as
\begin{equation}\frac{\rmd}{\rmd s}h(A(k),c(s))=h(A(k),c(s))A(k,c(s)),\end{equation}
and solutions to this equation will be 
\begin{equation}\label{eq:holo_solution}
h(A(k),c(s))=\mathcal{P}\exp\Biggl(\int_c A(k,c(s))\Biggr).
\end{equation}

We are going to define our exponential map as the standard matrix exponential map on a specific representation of $\mathfrak{su}(2)$ on each edge. This will require the exponential map be formally different on each edge. This is exactly what is usually done in LQG since the spin on each edge labels a specific representation of the gauge group.

An irreducible representation $\pi_\lambda: \mathfrak{g} \to \mathfrak{gl}_\lambda(V)$ is a Lie algebra homomorphism on the space $\mathfrak{gl}_{\lambda}(V)$ of $\lambda+1$-dimensional anti-Hermitian, traceless matrices over a vector field $V$, where the Lie bracket is given by the commutator of matrices. We also have the standard injection $i:\mathfrak{g}\to \mathcal{U}(\mathfrak{g})$. By the universality property, there exists a Lie algebra homomorphism $\theta_\lambda:\mathcal{U}(\mathfrak{g})\to \mathfrak{gl}_\lambda(V)$ such that the diagram commutes, $\theta_\lambda\circ i=\pi_\lambda$.

Now we have the standard exponential map $exp:\mathfrak{gl}_\lambda(V)\to GL_\lambda(V)$, defined by 
\begin{equation}\label{eq:exp}
exp(a):=1+a+\frac{a^2}{2!}+\frac{a^3}{3!}+...,~a\in \mathfrak{gl}_\lambda(V).
\end{equation}
We can construct an exponential map $exp_\lambda:\mathcal{U}(\mathfrak{g})\to GL_\lambda(V)$ by composition (see also figure \ref{fig:exp_diagram}):
\begin{equation}\label{eq:exp_l}
exp_\lambda(u):=exp\circ\theta_\lambda (u)~\forall u\in\mathcal{U}(\mathfrak{g}).
\end{equation}
This map is dependent on the representation $\pi_\lambda$ that we choose. Although it would be preferable to define this exponential map in a representation-independent way, it is not at all clear how to do this. A formal exponential map can be defined on the free Lie algebra of a set \cite{Serre_LieBook}, but the image of this map is inside $\mathcal{U}(\mathfrak{g})$ rather than a Lie group on which we can define cylindrical functions in the usual way. This construction can be extended by using the language of model theory and embedding the universal enveloping algebra into an ultraproduct of matrix Lie algebras \cite{L'Innocente}. This gives an exponential map with image in an ultraproduct of representations of the Lie group. In fact, the family of exponential maps we have defined above is inspired by those used on $\mathcal{U}(\mathfrak{sl}(2,\C ))$ in \cite{L'Innocente}, which they show can be restricted to $\mathfrak{su}(2)$. We might be able to define cylindrical functions on an ultraproduct of representations of $SU(2)$, but we view this construction as not explicit enough for our use. In our approach the exponential takes values in the locally compact group $GL_\lambda(V)$, and we can still use the Haar measure to define the inner product of cylindrical functions on this group. 

There is an alternative way of defining an exponential map that would work for our purposes, without having to invoke the general model theory result of \cite{L'Innocente}. This is to consider the standard angular momentum representation $\oplus_\ell V_\ell$ of $\mathfrak{su}(2)$ and identify elements in the universal enveloping algebra with the ``polynomial angular momenta" (higher order differential operators). Then the corresponding exponentials are the unitary operators associated to the polynomial angular momenta. One can truncate to a certain order of $\ell$ and obtain a locally compact group. If one restricts to only the first order differential operators, one again obtains $SU(2)$ and the usual exponential map. This approach is slightly simpler, but we consider the truncation of the spin labels to be undesirable in the present case.

The compactness of the holonomy group $SU(2)$ is a subtle, yet important element of the usual LQG theory. Formulating classical gravity with an $SU(2)$ connection rather than its complexification $SL(2,\C )$ is a result of taking the Immirzi parameter in (\ref{eq:connection}) to  be real. There have been attempts to construct LQG in the non-compact context \cite{Freidel-Levine}, including a recent proposal using projective techniques \cite{Okolow-2009}, but no one has been completely successful in implementing diffeomorphism invariance on such a Hilbert space \cite{Okolow}. We avoid this problem in this study by working with a locally compact group and basing our gauge invariance not on the measure, but on the states and operators of the theory.

\begin{figure}
\begin{center}\leavevmode
\begin{xy}
(0,0)*+{\mathcal{U}(\mathfrak{g})}="U"; (0,-25)*+{\mathfrak{g}}="su";(25,-25)*+{\mathfrak{gl}_\lambda(V)}="su_l";(50,-25)*+{GL_{\lambda}(V)}="SU_l";
{\ar@{<-_{)}} "U";"su"}?*!/_2mm/{i};
{\ar@{->} "su";"su_l"}?*!/_2mm/{\pi_\lambda};
{\ar@{->} "su_l";"SU_l"}?*!/_2mm/{exp};
{\ar@{-->} "U";"su_l"}?*!/_2mm/{\theta_\lambda}
\end{xy}
\caption{The map $\theta_\lambda$ exists by the universality of $\mathcal{U}(\mathfrak{g})$, and the map $exp$ is the usual matrix exponential.}\label{fig:exp_diagram}
\end{center}
\end{figure}

Since the holonomy over a single element $\sigma_\mu\in S_n$ is naturally an element of a group, the full holonomy over an edge $e_I$ parametrized by $s$ will be the direct product
\begin{equation}
H(A,e_I(s))=\prod_{\mu \in |S_n|}h(A(\sigma_\mu),e_I(s)).
\end{equation}
This holonomy takes values in
\begin{equation}\tilde{\mathcal{G}}_I:=Maps(S_n,GL_{\lambda_I}(V)))=GL_{\lambda_I}(V)^{|S_n|},\end{equation}
where $|S_n|$ is the index of $S_n$ and $\lambda_I+1=2\j_I+1$ is the dimension of the irreducible representation. $\tilde{\mathcal{G}}_I$ is locally compact by virtue of the finite index of $S_n$. This group has an action on it by $S_n$,
\begin{equation}\label{eq:Action_Finite}
\sigma'(\gamma_\sigma)=\gamma_{\sigma\sigma'},~\gamma_\sigma\in \tilde{\mathcal{G}}_I,~\sigma,\sigma'\in S_n.
\end{equation}
However, we have still not taken into account the fact that the holonomy has the same symmetry under deck transformations that the metric does. This can be determined directly by counting the identifications of the sheets by $\sigma_I$ over the edge $e_I$. Write the permutation as product of disjoint cycles,
\begin{equation}
\sigma_I=\beta_1...\beta_{m_I},
\end{equation}
where every element moved by $\beta_i$ is fixed by $\beta_j$. This decomposition is unique up to the ordering of the disjoint cycles. Each $\beta_i$ identifies sheets of the cover with one another, so the number of disjoint cycles $m_I$ is the number of identifications. If we mod out by this action we will get $|S_{m_I}|$ copies of $GL_{\lambda_I}$, where $m_I$ is the number of preimages $p^{-1}(e_I)$. Thus, the holonomy on the edge $e_I$ actually takes values in
\begin{equation}\mathcal{G}_I=\tilde{\mathcal{G}}_I/\sigma(e_I)=GL_{\lambda_I}(V)^{|S_{m_I}|}.\end{equation}
This also illustrates that the gauge symmetry here is $SU(2)\times K_I$, where $K_I$ is a finite group that interchanges sheets of the cover which are identified over $e_I$. We give a simple example of the holonomy on a topspin network in \ref{ap:ex}.

With these specifications we can solve the holonomy equation (\ref{eq:holo}), and we take the holonomy $h(A,c(s))$ to be the coordinate variables. It is easy to see this holonomy has the usual local transformation properties under the gauge transformation $\lambda: \Sigma \to SU(2)\times S_n$, because the form of the solution to (\ref{eq:holo}) is unchanged. Specifically, under the gauge transformation $A(x)^{\lambda}= -\rmd\lambda(x) \lambda(x)^{-1}+\lambda(x) A(x) \lambda(x)^{-1}$, the holonomy transforms like

\begin{equation}\label{eq:holo_transform}
h(A^\lambda,c(1))=\lambda(c(0))h(A,c(1))\lambda(c(1))^{-1}.
\end{equation}

For the conjugate variable to the holonomy, following the usual approach we will need to smear $E_a^{j\mu}$ with an $\mathcal{A}$-valued test function over a surface of dimension 2 \cite{TTbook,AshIII}. Thus the classical algebra on which our quantum theory will be based will be given by the holonomy (\ref{eq:holo_solution}) and the fluxes, which are integrated over a 2-surface $S$:

\begin{equation}\label{eq:E_flux}
E(S,f):=\int_S f_{j\alpha}(*E)^{j\alpha},
\end{equation}
where $(*E)^{j\alpha}=\epsilon^c_{~ab}E_c^{j\alpha}\rmd x^a\wedge \rmd x^b$. 

\section{Cylindrical Functions and the Hilbert Space}\label{s:cyl}
We will now construct the cylindrical functions in this new framework. We will closely follow the approach of \cite{TTbook}, just having to deal with  our holonomies living in the locally compact $\mathcal{G}_I$ rather than $SU(2)$. Before doing this however, we will discuss how we can combine the usual notion of a spin network on a spatial section with the topological information given by a branched cover of $\mathbb{S}^3$ so that the two structures are mutually compatible.

See figure \ref{fig:branch}, which shows the spatial section $\Sigma$ as an embedding in our spacetime manifold $M$ and the graph $\overline\Gamma\subset \Sigma$ on which we have some spin network labels $(\bar{j},\bar \iota)$. We know that this spatial manifold $\Sigma$ can be described as a branched cover of $\mathbb{S}^3$ by specifying an embedded graph $\tilde{\Gamma}\subset \mathbb{S}^3$ and a coloring of the edges $e_I$ of the graph with permutations $\sigma_I\in S_n$.

\begin{figure}
\begin{center}\leavevmode
\begin{xy}
(0,0)*+{M}="m"; (15,0)*+{\Sigma}="s";(22,0)*{\supset\overline{\Gamma}};
(15,-15)*+{\mathbb{S}^3}="s3"; (37,-15)*{\supset\tilde{\Gamma}\sim\Gamma=p(\overline{\Gamma})\bigcup\tilde{\Gamma}};
{\ar@<-.5ex>@{<-^{)}} "m";"s"};
{\ar "s";"s3"}?*!/_2mm/{p}
\end{xy}
\caption{Adding topological labels to a spin network to represent the spatial section as a branched cover. In this diagram $\sim$ means covering move equivalent.}\label{fig:branch}
\end{center}
\end{figure}

The image of the spin network in the 3-sphere is $p(\overline\Gamma)$, and since the decorated branch locus $\tilde{\Gamma}$ simply encodes the topological information for $\Sigma$, we can adjoin the two graphs $\Gamma=p(\overline\Gamma)\cup \tilde{\Gamma}$ and make it into a topspin network by adding trivial representation labels $(\tilde{j},\tilde\iota)=(0,Id)$ to $\tilde{\Gamma}$ and trivial permutation labels $\bar\sigma=(1)$ to $\overline\Gamma$. Thus the topological information about $\Sigma$ is preserved and the spin network information is contained completely in the trivial part of the branch locus.

To define the Hilbert space of cylindrical functions on topspin networks, we first choose a specific planar diagram $D(\Gamma)$ for an oriented graph $\Gamma\subset \mathbb{S}^3$. This diagram contains \textit{nodes}, which are either vertices of $\Gamma$ or oriented crossings of the diagram, and \textit{arcs}, which are either \textit{edges} of $\Gamma$ (with two endpoints at vertices) or curves that end on at least one crossing. Hereafter we will suppress writing $\Gamma$ in $D(\Gamma)$ for notational simplicity. We have a groupoid $l(D)$ of the diagram $D$ whose objects are nodes, and whose morphisms are the $L$ oriented arcs of $D$. Subgroupoids $l'(D')\subset l(D)$ represent subgraphs $\Gamma ' \subset\Gamma$ with the restricted planar diagram $D'=D|_{\Gamma'}$. Since the holonomy on each arc $H(e_I)$ for $I=1,\ldots,L$ is a map to a different matrix group $\mathcal{G}_I=GL_{\lambda_I}(V)^{|S_m|}$, we will define the direct product
\begin{equation}\label{eq:gauge_group}
\mathcal{G}:=\mathcal{G}_1\times \mathcal{G}_2\times \ldots \times \mathcal{G}_L.
\end{equation}
There are homomorphisms $x_{l(D)}:l(D)\to \mathcal{G}$ which are defined on the $J$th edge to be
\begin{equation}x_{l(D)}(e_J)=1\times\ldots 1\times H(e_J)\times 1\ldots \times 1.\end{equation}
The set of all such homomorphisms is $X_{l(D)}=Hom(l(D),\mathcal{G})$. These are homomorphisms in the sense that they preserve the structure of the groupoid. If the edges $e_I,~e_J$ can be composed then the resulting group elements can be multiplied in the direct product:
\begin{eqnarray*}
x_{l(D)}(e_I e_J)&=(1\times ... \times H(e_I) \times ... \times)\cdot(1 \times ...\times H(e_J) \times ...\times 1)\\
&=(1 \times ... H(e_I) \times H(e_J) \times ... \times 1)\\
&=x_{l(D)}(e_I)\cdot x_{l(D)}(e_J).
\end{eqnarray*}

We can now follow the usual approach taken in \cite{TTbook}. We have projections 
\begin{equation}\label{eq:projections}
p_{ll'}:X_{l(D)}\to X_{l'(D')},~\forall l'(D')\subset l(D),
\end{equation}
which are restrictions of the homomorphism $x_{l(D)}$ to the subgroupoid $l'(D')$. This gives us a projective family $(X_{l(D)},p_{ll'})$. Since we have a bijection
\begin{equation}\label{eq:bijection}
\rho_{D}:X_{l(D)}\to \mathcal{G},\quad \rho_D(x_{l(D)}(e_I))=1\times...H(e_I) ...\times 1
\end{equation}
and $\mathcal{G}$ is a locally compact group, we can equip $X_{l(D)}$ with a locally compact topology. We can now form the direct product of these spaces associated to each groupoid $l(D)$,
\begin{equation}X(D)=\prod_{l(D)} X_{l(D)},\end{equation}
and equip it with the product topology. This space depends on the choice of diagram for each graph.

Still following \cite{TTbook}, we define our cylindrical functions as
\begin{equation}f\in Cyl'(X(D)):=\bigcup_{l(D)}C_0(X_{l(D)}),\end{equation}
where $C_0(X_{l(D)})$ are functions of finite support on $X_{l(D)}$ since $\mathcal{G}$ is only locally compact. If we restrict to first order terms in $\mathcal{U}(\mathfrak{g})$, we can recover using continuous functions here although we would lose the algebraic structure of the fields. There is an equivalence relation
\begin{equation}f\sim f'\mbox{ if }p^*_{l''l}f=p^*_{l''l'}f'~\forall l(D),l'(D)\subset l''(D),\end{equation}
and we define $Cyl(X(D)):=Cyl'(X(D))/ \sim$. As discussed in \cite{TTbook}, these cylindrical functions form a *-algebra with the sup norm
\begin{equation}\Vert f\Vert:=sup_{x_l(D)}\vert f_{l(D)}(x_{l(D)})\vert,~f_{l(D)}\in C_0(X_{l(D)}).\end{equation}
We illustrate the general form for the cylindrical functions over a simple topspin network in \ref{ap:ex}.

The cylindrical functions are still defined on an arbitrary diagram $D$, but we can remove the dependence on this diagram in the following manner. We can write 
elements of $X_{l(D)}= Hom(l(D),Maps(S_m,GL_\lambda(V)))$ in the form 
$x_l(e_1\ldots e_m)=g_{e_1}(\sigma)\cdots g_{e_m}(\sigma)$, with $\sigma \in S_m$.
We want to define versions of $X_{l(D)}$ and $X(D)$ that will no longer 
depend on the planar diagram $D=D(\Gamma)$, but only on (the ambient 
isotopy class of) the embedded graph $\Gamma$. To this end we consider
instead of $X_{l(D)}$ the subset $Y_{l(D)}\subset X_{l(D)}$ given by only those
$x_l \in Hom(l(D),Maps(S_m,GL_\lambda(V)))$ of the form $x_l (e_1\ldots e_m)=g_{e_1}(\sigma)\cdots g_{e_m}(\sigma)$ with
$$ g_{e_I}(\sigma) = \left\{ \begin{array}{ll} 0 & \sigma\neq \sigma(e_I) \\
g(\sigma(e_I)) & \sigma= \sigma(e_I), \end{array}\right. $$
where the $\sigma(e_I)$ are the permutations associated to the oriented 
arcs $e_I$ of the planar diagram $D$ in the Wirtinger presentation of
$\pi_1( \mathbb{S}^3 \smallsetminus \Gamma)$ associated to the planar
diagram $D=D(\Gamma)$, and $g: S_m \to GL_\lambda(V)$ is a group homomorphism.
This restriction has the effect of selecting only those $x_l(e_1,\ldots,e_m)=g(\sigma(e_1))\cdots g(\sigma(e_m))$, where the $\sigma(e_I)$ satisfy the Wirtinger relations. These $x_l$ can then be viewed as groupoid homomorphisms from $l(D)$ to $\pi_1( \mathbb{S}^3 \smallsetminus \Gamma)$ composed with a group homomorphism to $GL_\lambda(V)$. Because the Wirtinger relations imply the invariance with
respect to Reidemeister moves, we can then identify $Y_{l(D)}$ with $Y_l$, independently of the choice of a planar diagram. We can then define the total space $Y=\prod_{l}Y_l$ as we did before, but with $Y$ now independent of the choice of diagram for each $l$.

The next step will be to define a measure for the inner product on our Hilbert space. Since the cylindrical functions here are defined on finite products of locally compact topological groups $GL_\lambda(V)$, the construction again follows from \cite{TTbook} until we come to the issue of gauge invariance.

We can identity the space $X_l$ with the group $\mathcal{G}$ via the bijection (\ref{eq:bijection}). When passing to $Y_l$ as above, we can similarly identify it with a subset of $\mathcal{G}$, where instead of all set theoretic maps $ S_n \times E(\Gamma) \to GL_\lambda(V) $ for edges $E(\Gamma)$ of $\Gamma$, we only consider those maps that satisfy the conditions described above, \textit{i.e.} that implement the Wirtinger relations.

The measure on each $X_l$ will be defined in terms of the Haar measure $\mu_{H}$ on each copy of $GL_{\lambda_I}(V)$. The measure on each $\mathcal{G}_I$ for edge $e_I$ will be $\rmd \tilde{\mu}(H(e_I))=\prod_{k\in |S_{m_I}|}\rmd\mu_H(h(k,e_I))$ for the holonomy $h(k,e_I)$ in the $k$th copy of $GL_{\lambda_I}(V)$. We also need to pull the cylindrical functions back to the group $\mathcal{G}$, so we define
\begin{equation}f_l(x_l):=(\rho^*_l\Psi_l)(x_l),~x_l\in X_l,\end{equation}
where $\Psi_l\in C_0(\mathcal{G})$. The measure on $\mathcal{G}$ for a specific subgroupoid $l$ is
\begin{equation}\label{eq:measure}
\mu_l(f_l)=\int_{\mathcal{G}}\Biggl( \prod _{I=1}^{L}\rmd\tilde{\mu}_H (H(e_I))\Biggr)\Psi_l(H(e_1)H(e_2)...H(e_L)).
\end{equation}
We can then take the product measure $\mu=\prod_{l}\mu_l$ on all the possible subgroupoids for our graph and define the Hilbert space of cylindrical functions as
\begin{equation}\mathcal{K}_D:=L_2(\overline{Cyl(X(D))},\rmd\mu).\end{equation}

The measure constructed in this way on $X_l$ induces a measure on $Y_l$ of
non-zero total mass, and we obtain corresponding Hilbert spaces of cylindrical functions
\begin{equation}\mathcal{K}:=L_2(\overline{Cyl(Y)},\rmd\mu).\end{equation}
In \cite{TTbook} the measure $\mu$ is constructed by showing that the family of measures $\mu_l$ can be consistently pulled back to $\mu$ via the projections (\ref{eq:projections}). This should work in our case as well, but we do not need the extra technical considerations.

The above measure $\mu$ will not necessarily be gauge invariant; the gauge invariance in the traditional approach is related to the left-right invariance of the Haar measure on $SU(2)$ \cite{Rovelli,TTbook}. The holonomy transforms under (\ref{eq:holo_transform}), so the cylindrical functions will transform as
\begin{equation}
\fl\Psi(h_1h_2...h_L)\mapsto \Psi(\lambda(x^{1}_f) h_{1}\lambda^{-1}(x^{1}_i)\lambda(x^{2}_f) h_{2}\lambda^{-1}(x^{2}_i)...\lambda(x^{L}_f) h_{L}\lambda^{-1}(x^{L}_i)).
\end{equation}
It follows immediately that the inner product on $\mathcal{K}$ is invariant if the Haar measure is left-right invariant. This would not be true in the current locally compact case, but since there is always a gauge invariant subspace $\mathcal{K}_0$ of the Hilbert space (the spin network states), this measure will be gauge invariant on those states. We will always restrict to the spin network states $\mathcal{K}_0\subset \mathcal{K}$ when using this inner product.

Our cylindrical functions now have an extra label $\vec{\sigma}$ to denote the set of permutation labels on the graph. An interesting difference from the usual case is that since we want to consider all possible graphs (up to ambient isotopy classes, see below), the Hilbert space now contains cylindrical functions on graphs which describe \textit{all} compact oriented three-manifolds $\Sigma$, rather than spatial sections of some Lorentzian four-manifold. We would then expect the classical limit to somehow select only those networks which lead to spatial sections of a spacetime, or set cylindrical functions on graphs which do not describe such sections to vanish. Of course, the classical limit is not yet well-defined in the loop framework so we can not confirm this intuition. In the case of Riemannian four-manifolds we should have no problem taking any codimension 1 submanifold we like and representing it as a covering space.

The Hilbert space of usual LQG can be made separable by considering equivalence classes of $\mathcal{K}_0$ under extended diffeomorphisms $\phi\in Diff^*$, which are smooth everywhere except at a finite number of points. Then the Hilbert space $\mathcal{K}_{Diff^*}=\mathcal{K}_0/\sim$ has a basis $|K^*,\mathcal{C}\rangle$ labeled by the knot class $K^*$ and coloring $\mathcal{C}$ of the intertwiners and arcs. These states are usually referred to as s-knot states. It is well known from knot theory that the set of ambient isotopy classes of loops (knots without intersections) is a countable set, and it can be shown that under the extended diffeomorphism group knots with intersections remain in the class of the underlying loop with vertices removed \cite{Fairbairn-Rovelli}. Therefore the s-knot states form a countable basis for the Hilbert space $\mathcal{H}_{Diff*}$, and in the usual theory the (extended) diffeomorphism invariance is what separates the Hilbert space.

In principle we would like our Hilbert spaces to also be separable, and it turns out they are. However, due to the addition of the topological labels, we cannot use the same set of graphs as in the usual theory. It can be shown that two topologically different three-manifolds can be realized as branched covers over the same graph (we give such an example in \ref{ap:knots}). In fact, the Hilden-Montesinos theorem tells us that this can even be done by restricting to special classes of links \cite{Knots}. So we will get a separable Hilbert space if we consider the discrete set given by the equivalence class of extended diffeomorphisms on graphs and (for a fixed order $n$ of the covering) a finite number of orientation, spin, and topological labels (since the topological labels will differentiate the topology). There is still the issue of varying the order of the covering, but that just adds a discrete label $n$, so the resulting Hilbert space is still separable. One needs to take into account the equivalence relation given by stabilization (adding trivial coverings) discussed in \cite{DMA}. Notice that we can also appeal to the Hilden-Montesinos result to specialize to three-fold covers. The advantage of allowing more general coverings and of working with embedded graphs instead of restricting to knots and links is that one can then implement certain composition operations, as described in \cite{DMA}.

The diffeomorphism invariance of this theory should be investigated in more detail by looking at diffeomorphims (or extended diffeomorphisms) on the spatial sections $\Sigma$, which are the physical gauge transformations. For a spin network $(\bar{\Gamma},\bar{j},\bar{\iota})$, a spatial diffeomorphism $\phi:\Sigma\to \Sigma'$ can drag the graph around the manifold as well as change the orientation and coloring of the links. As discussed in \cite{Rovelli}, it is possible to project to the diffeomorphism-invariant states such that if the graph changes, the corresponding cylindrical functions are orthogonal. Then the action of a diffeomorphism on an invariant cylindrical function is equivalent to the action of a finite group $G_{\overline\Gamma}$ which permutes the labels and orientations of the underlying spin network function. Of course, since $\phi$ is supposed to be a diffeomorphism, the topspin network which is the branch locus for $\Sigma$ must be equivalent under covering moves to the topspin network associated with $\Sigma'$. Thus determining the diffeomorphism invariance of the cylindrical functions would require checking the compatibility of the cylindrical functions under covering moves. We leave this technical result for a later study. 

\section{Operators}\label{s:operators}
To define the operators in this theory we will closely follow the description by \cite{Ash_Review,Rovelli,TTbook,AshI}. One must regularize these by partitioning the surface $S\subset \mathbb{S}^3$ over which the phase space variables are smeared into $N$ smaller surfaces, and then take $N\to \infty$. This surface lives in the base space, and is the image of the physical surface $\tilde{S}$ under the covering map, $p(\tilde{S})=S$. Since we are not altering the structure of the spin networks (only adding extra labels to them) and $S$ acts just like the usual surfaces in LQG, there should be no change in the regularization procedure. Therefore we will not discuss it here.

\subsection{The Area Operator}\label{s:Area_Operator}
The classical operators in our theory are the holonomies $h(A,c)$ and the flux vector fields $E(S,f)$. Let $S$ be an oriented, embedded, open, compactly supported surface with embedding $X:U\hookrightarrow S$ from open set $U\subset \mathbb{R}^2$. The (classical) area of $S$ would be

\begin{equation}\label{eq:classical_area}
A(S)=\int_S \sqrt{n_a E^{ai}(S,f)n_bE^{bj}(S,f)\kappa_{ij}}\rmd ^2x,
\end{equation}
where
\begin{equation}n_a:=\epsilon_{abc}\frac{\partial X^b}{\partial x^1}\frac{\partial X^c}{\partial x^2},\end{equation}
$\kappa_{ij}$ is the Killing form on $\mathfrak{su}(2)$ and $(x_1,x_2)$ are coordinates on $S$. We would like to use this same approach in our case, but there will be one important difference. In (\ref{eq:classical_area}), since $E^a\in\mathfrak{su}(2)$ the area operator will be proportional to $\sqrt{\kappa_{ij}\tau^i\tau^j}$, the (quadratic) Casimir operator on the Lie algebra $\mathfrak{su}(2)$. In other words, the Casimir is the trace of two first-degree differential operators. In our case, since $E^a_{\mu}\in \mathcal{U}(\mathfrak{su}(2))$, the inner product will have to be on differential operators of arbitrary degree. Thus, to extend this result and retain gauge invariance we need a field $E^a_{\mu}$ that takes values in the center of $\mathcal{U}(\mathfrak{su}(2))$. This does not in fact change the situation much from the usual case, it just requires a bit more care. For instance, if we have a single second-order operator $\hat{O}_a\tau$ with basis element $\tau=\kappa_{ij}\sigma^{i}\sigma^j$ of $\mathcal{U}(\mathfrak{su}(2))$, then
\begin{equation}\hat{O}=\int_{\mathbb{R}}\hat{O}_a \rmd x^a\end{equation}
would be gauge-invariant (since it is proportional to an element in the center). To clarify this point, we will use the natural grading on $\mathcal{U}(\mathfrak{su}(2))$,
\begin{equation}\label{eq:grading}
\gamma (\exp_{\lambda_1} (u_1)...\exp_{\lambda_k} (u_k))=\gamma^{n_1+n_2+...+n_k}(\exp_{\lambda_1} (u_1)...\exp_{\lambda_k} (u_k)),
\end{equation}
for $u_i$ a polynomial in the universal enveloping algebra with degree $n_i$ in the representation $\lambda_i$. Then we can define a family of area operators $\leftexp{(k)}{A(S)}$, where the first order one will be exactly as (\ref{eq:classical_area}) with the restriction $E_a^{i\mu}\to \leftexp{(1)}{E_a^{i\mu}}$. To calculate this we split up the surface $S=\bigcup _{I\in N}S_I$ into $N$ surfaces, each intersecting 0 or 1 edges. We also assume that the vertices of the network lie outside $S$. The first order area operator then becomes
\begin{equation}\label{eq:area_N}
[\leftexp{(1)}{A(S)}]_N:=\sum_{I=1}^N \sqrt{\leftexp{(1)}{E^{i\mu}_a}(S_I)\leftexp{(1)}{E^{aj\nu}}(S_I)\kappa_{ij}\kappa_{\mu\nu}}\rmd x^1\wedge \rmd x^2.
\end{equation}
In general we should replace the Killing form with an appropriate extension to $\mathcal{U}(\mathfrak{su}(2))$, but since we have restricted ourselves to first order operators we can use the usual Killing form for now. The inner product $\kappa_{\mu\nu}$ on $\C S_n$ implements (\ref{eq:CS_product}) on the basis elements 
\begin{equation}(\sigma^{\mu},\sigma^{\nu})=\frac{n^2}{|S_n|}\sigma^{\mu}\sigma^{\nu}\kappa_{\nu\mu}=\frac{n^2}{|S_n|}\sum_{\sigma^{\mu},\sigma^{\nu}\in S_n}\delta_{\mu\nu}=n^2.\end{equation}

The calculation of the area operator exactly follows the usual one (see \cite{Ash_Review,TTbook,AshI}), so we will simply describe the result. Assuming that the vertices of the topspin network do not lie within $S$ (this is the \textit{restricted spectrum}), the Killing form on $\mathfrak{su}(2)$ gives the usual $j_I(j_I+1)$ (since we have restricted to the first-order part). However, due to the symmetry on the holonomy elements under the action of (\ref{eq:Action_Finite}) from edge $e_I$, the inner product (\ref{eq:CS_product}) actually takes place in $S_{m_I}$ with $m_I\leq n$ as discussed in \S\ref{s:2.2}. Our basis elements would then be in an $m_I$-dimensional representation $\rho:S_{m_I}\to GL(\mathbb{C}^{m_I})$. Thus, we find that for the restricted spectrum, the area eigenvalues depend on how the cover is stitched together above each edge. The overall result is
\begin{equation}\label{eq:Area}
\leftexp{(1)}{A(S)}=4\pi l_P^2\beta\sum_{I=1}^N\sqrt{j_I(j_I+1)\cdot m_I^2},
\end{equation}
where we now just sum over the $N$ intersections of $S$ with $\Gamma$.

In the above, we restricted the fields to be first order, which makes the area operator (\ref{eq:Area}) gauge invariant. There are also higher-order gauge invariant operators $\leftexp{(k)}{A(S)}$ which lie in the center of $\mathcal{U}(\mathfrak{su}(2))$. The structure of these can be clarified using the fact that there is a bijection between the universal enveloping algebra and the symmetric algebra $\odot {\mathfrak{g}}$ \cite{Dixmier}. We can use this bijection to write a basis of $\mathcal{U}(\mathfrak{g})$ in symmetric polynomials of the generators $T_j$ of $\mathfrak{g}$,
\begin{equation}P_W(\tau)=\sum_{k=0}^\infty \Pi^{i_1...i_k}T_{i_1}...T_{i_k}.\end{equation}
These are called \textit{Weyl-ordered polynomials}; each of tensors $\Pi^{i_1...i_k}$ are completely symmetric. Such a polynomial will be in the center of the universal enveloping algebra if the adjoint action 
\begin{equation}Ad(T_k)P_W(T)=[T_k,P_W(T)]\end{equation}
vanishes for all $T_k\in \mathfrak{g}$. By writing out the adjoint action on each term in $P_W(T)$ and using the structure functions defined by $[T_i,T_k]=f_{ik}^{~~j}T_j$, it is easy to see that $P_W(T)$ will be in the center of $\mathcal{U}(\mathfrak{g})$ if
\begin{equation}\label{eq:center}
f_{kl}^{~~i_1}\Pi^{li_2...i_n}+...+f_{kl}^{~~i_2}\Pi^{i_1...i_{n-1}l}=0.
\end{equation}
Given a tensor $\overline{\Pi}^{i_1...i_k}$ that satisfies this, we can write a general order $k$ Casimir element as
\begin{equation}\mathcal{C}_k=\overline{\Pi}^{i_1...i_k}T_{i_1}...T_{i_k}.\end{equation}
Specializing to $k=2$, we find we need a rank-2 symmetric tensor which is invariant under the adjoint action. These properties are satisfied by the Killing form $\kappa_{ij}$, which is why it was used in (\ref{eq:classical_area}). This makes explicit what we discussed at the end of \S\ref{s:variables}; $\mathcal{U}(\mathfrak{g})$ is naturally the algebra of observables which take values in a Lie algebra $\mathfrak{g}$. We simply allowed our fields to take values in this algebra \textit{a priori}.

We choose to call the $k$th-order area operator the one which contains fields which are order $k$ polynomials in the generators of $\mathfrak{su}(2)$. In the basis given by the Weyl-ordered polynomials, our fields will be a sum of all possible order-$k$ polynomials in the symmetric $\mathfrak{su}(2)$-valued indices:
\begin{equation}
E=\sum_{k=0}^{\infty}\leftexp{(k)}{E},~ \leftexp{(k)}{E}=(E^{\mu}_a)^{i_1...i_k}T_{i_1}...T_{i_k}\otimes \sigma_{\mu}dx^a.
\end{equation}
We define the order $k$ area operator then as (ignoring regularization),
\begin{eqnarray}\label{eq:kA}
\leftexp{(k)}{A(S)}&=\sum_{I=1}^{N} \sqrt{(E^\mu_a)^{i_1...i_k}(S_I)(E^{a\nu})^{j_1...j_k}(S_I)\overline{\Pi}_{i_1...i_kj_1...j_k}\kappa_{\mu\nu}}\\
&=\sum_{I=1}^{N} \sqrt{\mathcal{C}_k(S_I)n_I^2}.
\end{eqnarray}
This gives an infinite family of gauge invariant area operators. Such operators would exist in the usual LQG formalism as polynomials in the fields $E^j_a$, but in that case it is not clear how operators like $E^jE^kE^l$ would be related to the metric. Here our metric is an inner product of fields $\eta_{ij}E^iE^j$ taking values in $\mathcal{U}(\mathfrak{su}(2))$, so the connection between the classical area operator (\ref{eq:classical_area}) and the higher-order gauge invariant operators (\ref{eq:kA}) is more direct. The physical meaning of these higher-order gauge invariant operators is unknown; we just note their existence.

How is the area of a surface $S\subset \mathbb{S}^3$ related to the physical surface $\tilde{S}=p^{-1}(S)\subset \Sigma_t$? It is easy to see that $A(\tilde{S})=\leftexp{(1)}{A(S)}$; this is because the field $E_a^{i\mu}(S_I)$ on an edge $e_I\in\Gamma \hookrightarrow \mathbb{S}^3$ can also be written as a field on the (equivalence class of the) $i$th preimage $p^{-1}(e_I)\in \Sigma_t$. The first-order area operator can be evaluated on the surface $\tilde{S}$, 
\begin{equation}\label{eq:tildearea}
\leftexp{(1)}{A(\tilde{S})}=\sum_{I=1}^{N'} \sqrt{\leftexp{(1)}{E^{i}_a}(\tilde{S}_I)\leftexp{(1)}{E^{aj}}(\tilde{S}_I)\kappa_{ij}},
\end{equation}
where $E^i_a$ are the usual fields of LQG (\textit{i.e.} no topological labels) and $N'$ is the number of intersections $\tilde{S}\bigcap\tilde{\Gamma}$; (\ref{eq:tildearea}) is exactly the same as (\ref{eq:classical_area}). However, since we know the spatial manifold $\Sigma_t$ can be written as a branched cover with some of the edges projecting to a single edge in the base, some of the elements of this sum are going to have the same spin $j$. In other words, if $\Sigma_t$ can be represented as an $n$-fold branched covering space there might be sets of edges $e_{\alpha_i}$ that satisfy
\begin{equation}p^{-1}(e_K)=e_{\alpha_1}\cup e_{\alpha_2}\cup ... \cup e_{\alpha_{m_K}} \end{equation}
with $m_K\leq n$. These edges have the same spin label $j_K$, and would contribute $m_K\sqrt{j_K(j_K+1)}$ to the area operator. This result is exactly the same as (\ref{eq:Area}). Thus, although the surface $S$ is not the physical surface $\tilde{S}$, the result of the area operator $\leftexp{(1)}{A(S)}$ on an topspin network is the same as the usual area operator $A(\tilde{S})$ on the corresponding spin network.

The significance of the scaling in (\ref{eq:CS_product}) is now obvious; it was chosen so that the area operator on topspin networks would reproduce the usual one. Of course, this is what we expect since they are geometrically measuring the same thing; the number of intersections of the surface with the graph. The scaling in such an inner product is also arbitrary, unlike the Killing form which is defined in terms of the adjoint action of a Lie algebra. The first-order area operator (\ref{eq:Area}) can be said to ``track the topology'' in the sense that the set of integers $m_I$ comes from the branching, which depends on the topology. In principle, these integers would be dependent on the dynamics, which are governed by the Hamiltonian. We will discuss this further in the next two sections. Different values of $m_I$ do not necessarily correspond to topologically inequivalent three-topologies (nor should we expect them to), but this at least provides a framework for studying topology in LQG.

The area operator is not invariant under the geometric covering moves shown in figure \ref{fig:covering_moves}. This is essentially because the move $V_1$ allows one to split arcs of the graph, which can change the spectrum of an area operator which intersects them. This idea can be made precise; consider an arc of the topspin network of a 4-fold branched cover with spin 1 and the permutation label $(14)(23)\in S_4$. A surface $S$ which intersects with this arc contributes $\sqrt{1\cdot 2\cdot 2^2}=2\sqrt{2}$ to the area eigenvalue. Under the geometric covering move $V_1$ we can split this arc into two arcs, one with coloring $(\mathbf{1/2},(14))$ and one with $(\mathbf{1/2},(23))$. Since a single permutation identifies just two covers, we find that these two arcs contribute $2\sqrt{1/2\cdot 3/2\cdot 3^2}=3\sqrt{3}$. The geometric covering moves preserve the geometry of the branched cover, but the do \textit{not} preserve the surface $S$ or its area. Intuitively, by changing the number of identifications we are changing the number of self-intersections of the surface $\tilde{S}$, and we would not expect the area to be preserved by that process.

\subsection{The Hamiltonian}
Dynamics in LQG are determined by the action of the Hamiltonian (\ref{eq:Wheeler-DeWitt}). This encodes the Hamiltonian constraint of the canonical system. We will review the action of the Hamiltonian on spin networks and extend it to topspin networks in a natural way. We are going to continue to work with the real connection; for the Lorentzian theory you need a second term in the Hamiltonian \cite{TTbook}. First we recall some definitions which will be relevant for our discussions \cite{QSDII}.

\begin{definition}\label{def:extra_v}
An \textbf{extraordinary vertex} $v$ of a graph $\Gamma$ is a vertex that is at most trivalent and is the intersection of exactly two analytic curves $c,c'\in \Gamma$. In addition, $v$ must be an endpoint of $c$.
\end{definition}

\begin{definition}\label{def:extraordinary}
An \textbf{extraordinary edge} $e$ of a graph $\Gamma$ satisfies
\begin{itemize}
\item its endpoints $v_1,~v_2$ are extraordinary vertices,
\item the two curves $c,~c'$ which contain $v_1,~v_2$ but are not the curve $e$ must intersect in at least one vertex,
\item exactly one of the vertices of $c,~c'$ is called the \textbf{typical vertex}.
\end{itemize}
\end{definition}

\begin{figure}\begin{center}
\includegraphics[scale=0.5]{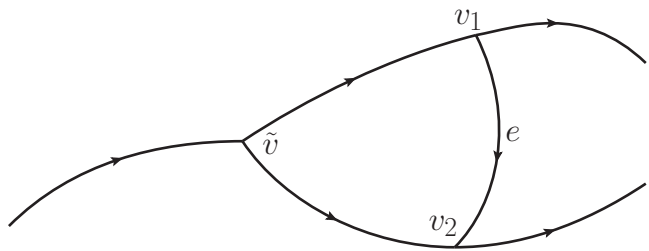}
\caption{An example showing the definitions \ref{def:extra_v} and \ref{def:extraordinary}. $v_1,~v_2$ are the extraordinary vertices, $e$ is the extraordinary edge, and $\tilde{v}$ is the typical vertex.}\label{fig:extra}
\end{center}\end{figure}
See Figure \ref{fig:extra} for an illustration of these definitions. There is an issue of choosing the typical vertex if $c,~c'$ intersect twice but we will not encounter this problem here - it is covered in \cite{QSDII}. These definitions work perfectly well for the topspin case as well since they do not refer to any spin labels. We can just take the graph to be $\Gamma\subset \mathbb{S}^3$ and use these definitions as they are.

\begin{figure}
\begin{center}
\includegraphics[scale=0.5]{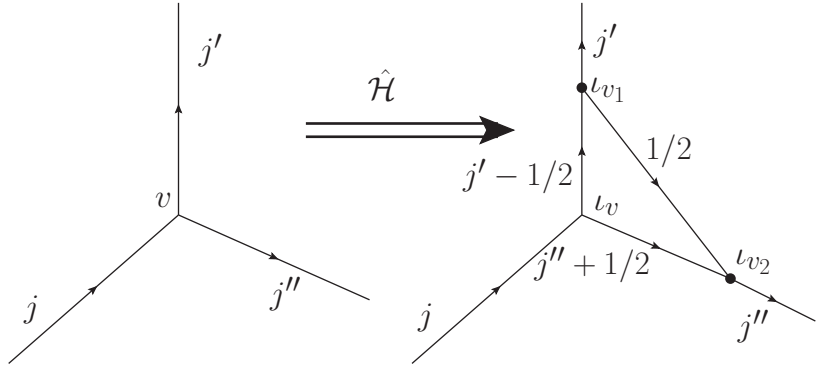}
\caption{The Hamiltonian acting on a vertex of a spin network state.}
\label{fig:H_spinnet}
\end{center}
\end{figure}

The definition of the Hamiltonian in LQG is subject to several ambiguities. We will choose the definition from \cite{Rovelli}, because it transforms covariantly under extended diffeomorphisms \cite{Fairbairn-Rovelli}. The action of the Hamiltonian operator on a spin network is completely specified by its action on vertices as shown in figure \ref{fig:H_spinnet} \cite{Rovelli,TTbook,QSDII,QSDI}. From the above definitions, we see that the Hamiltonian adds extraordinary edges to spin networks. For an $n$-valent typical vertex ($n> 2$), these edges can be added in $n!/[2!(n-2)!]$ ways. The Hamiltonian acts on all the vertices, so the full action of $\hat{\mathcal{H}}$ on a spin network is to produce a sum of spin networks with an extraordinary edge added at each vertex, chosen in different ways. This motivates the following definition:

\begin{definition}\label{def:spin-net}
A \textbf{spin-net} is a triple $\psi=(\Gamma,j,\iota)$, and the set of all such spin-nets we call $\mathcal{S}$. For a given spin-net, there exists a unique source spin-net $\psi_0$ which is constructed by removing all the extraordinary edges iteratively. Now define the sets $\mathcal{S}^n(\psi_0)$ by iteratively adding $n$ extraordinary edges in all topologically inequivalent ways, and set $\mathcal{S}^0(\psi_0)=\{\psi_0\}$.
\end{definition}
The details of this definition can be found in \cite{QSDII}. In that same work, it was found that different spin nets are totally disjoint:

\begin{theorem}\label{thm:disjoint}
\mbox{}
\begin{itemize}
\item $\mathcal{S}^n(\psi_0)\cap \mathcal{S}^{n'}(\psi'_0)=0$ if $n\neq n'$ or $\psi_0\neq \psi'_0$.
\item Every $\psi\in \mathcal{S}^n(\psi_0)$ has source $\psi_0$.
\end{itemize}
\end{theorem}
Since the Hamiltonian adds extraordinary vertices, one can use this theorem to find the kernel of the Hamiltonian and reduce the problem of finding physical states to one of linear algebra \cite{TTbook,QSDII}. 

To extend the action of the Hamiltonian to topspin networks, we consider it as a series of maps between spin networks $(\Gamma,j,\iota)\to (\Gamma',j',\iota')$ given by adding an extraordinary vertex to each. Specifically, consider the set of all maps
\begin{equation}f^n_v:\mathcal{S}^n(\psi_0)\to \mathcal{S}^{n+1}(\psi_0),\end{equation}
which adds an extraordinary edge to the vertex $v$ of a spin-net, which has source $\psi_0$. In principle this could be done in several different ways depending on the valence of the vertex $v$, but this consideration can easily be added in. This map changes the spin labeling $j$ to be compatible with new intertwiners $\iota_v,\iota_{v_1},\iota_{v_2}$ at $v$ and the two new vertices $v_1,~v_2$ at the ends of the extraordinary vertex. The action of the Hamiltonian on a spin network could be decomposed into a composition of such functions, one for each vertex.

This construction is naturally a category $\mathcal{C}_{\mathcal{H}}(\psi,G)$, with the spin networks as the objects and morphisms $f^n_v$. $\mathcal{H}$ denotes that this category is induced by the action of the Hamiltonian, and the $G$ represents the choice of structure group for the theory. We would like to extend this category to the case of topspin networks, which will require adding topological labels and enforcing the Wirtinger relations. 

\begin{figure}[b]
\begin{center}
\includegraphics[scale=0.5]{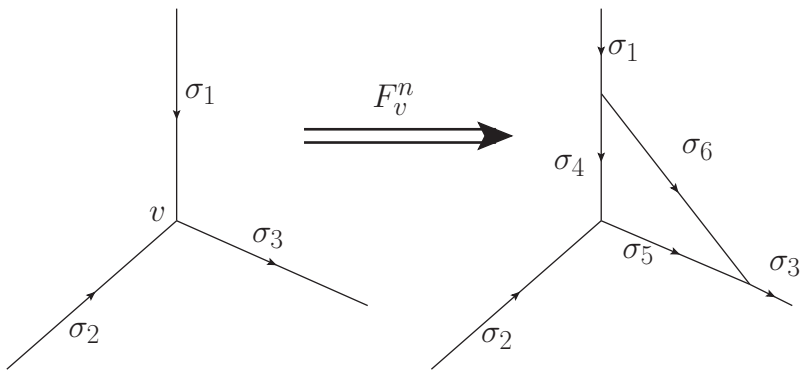}
\caption{The morphism $F_v^n$ acting on a vertex $v$ of a topspin state. The simplest non-trivial choice for $\sigma_6$ is a single transposition. The (suppressed) spin labels here are the same as in figure \ref{fig:H_spinnet}.}
\label{fig:H_topspin}
\end{center}
\end{figure}

If we include the permutation labels $\sigma_I$ on each edge $e_I$ in the topspin formalism, the original functions $f_v^n$ can be extended to include their action on the permutation labels. We thus define \textbf{topspin-nets} $\mathcal{T}^n(\Psi_0)$ with sources $\Psi_0$ in exactly the same manner as spin-nets except ensuring that the Wirtinger relations are satisfied for the permutation labels at the vertices. This can be easily done when reconstructing the topspin-net from the source as in definition \ref{def:spin-net}. The morphisms between topspin-nets are
\begin{equation}F^n_v:\mathcal{T}^n(\Psi_0)\to \mathcal{T}^{n+1}(\Psi_0).\end{equation}
The result of acting with this map on a topspin-net is shown in figure \ref{fig:H_topspin}. The relations between the representations on the new edges are given by the Wirtinger relations at the vertices. In the first case, we have just that
\begin{equation}\sigma_1\sigma_2\sigma_3^{-1}=1.\end{equation}
After the action of $F^n_v$ we have similar relations at the two new nodes, and the new representations are given by 
\begin{equation}\sigma_4=\sigma_1\sigma_6^{-1},\quad \sigma_5=\sigma_3\sigma_6^{-1}.\end{equation}
There is still some freedom in how one chooses $\sigma_6$, which is analogous to the freedom one has when choosing the representation $j=1/2$ in figure \ref{fig:H_spinnet}. In the spin network case the classical operator is unaffected by this, but the quantum operator may be \cite{Rovelli}. At this point in the analysis we have no guiding principal except the analogy to the spin network case, and therefore we choose the simplest non-trivial element for $\sigma_6$, a transposition of a pair of sheets. This extension of the category for spin networks we denote $\mathcal{C}_{\mathcal{H}}(\Psi,\mathcal{G})$. Again, we can define the Hamiltonian as the composition of the functions $F^n_v$, one for each vertex.

By looking at the structure of these two categories, it is easy to see that they are inequivalent; this means that the action of the Hamiltonian in the topspin framework is structurally different than in the usual case. The basic reason is the presence of the geometric covering moves (figure \ref{fig:covering_moves}) which impose an equivalence relation on the objects of $\mathcal{C}_{\mathcal{H}}(\Psi,\mathcal{G})$. These moves provide us with a way to ``remove extraordinary edges''. Such things are not possible in the usual LQG because there is not an appropriate notion of edge deletion. We now provide an example of this.

\begin{figure}
\begin{center}
\includegraphics[scale=0.4]{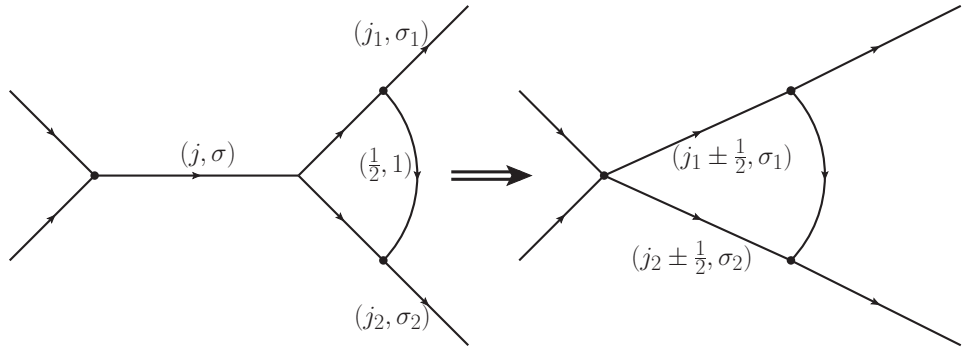}
\caption{The combined covering move V', where we have suppressed the intertwiners $\iota$.}
\label{fig:V'}
\end{center}\end{figure}

Consider the pair of vertices given in figure \ref{fig:V'}. We define the combined geometric covering move $V'=V_2\circ V_1$. This can change the first diagram to the second, provided the following is satisfied:
\begin{equation}
j=(j_1\pm 1/2)\otimes (j_2 \pm 1/2),~\sigma=\sigma_1\sigma_2.
\end{equation}
The result of this move is to destroy a vertex (the typical vertex associated to the extraordinary edge $(j,\sigma)=(\mathbf{1/2},(1))$ and make the vertex at the end of the edge that was split into the new typical vertex. Note that we must do this in such a way to preserve the properties of the extraordinary edge in definition \ref{def:extraordinary}. This move does not, by itself, destroy an extraordinary edge.

\begin{figure}
\begin{center}
\includegraphics[scale=0.4]{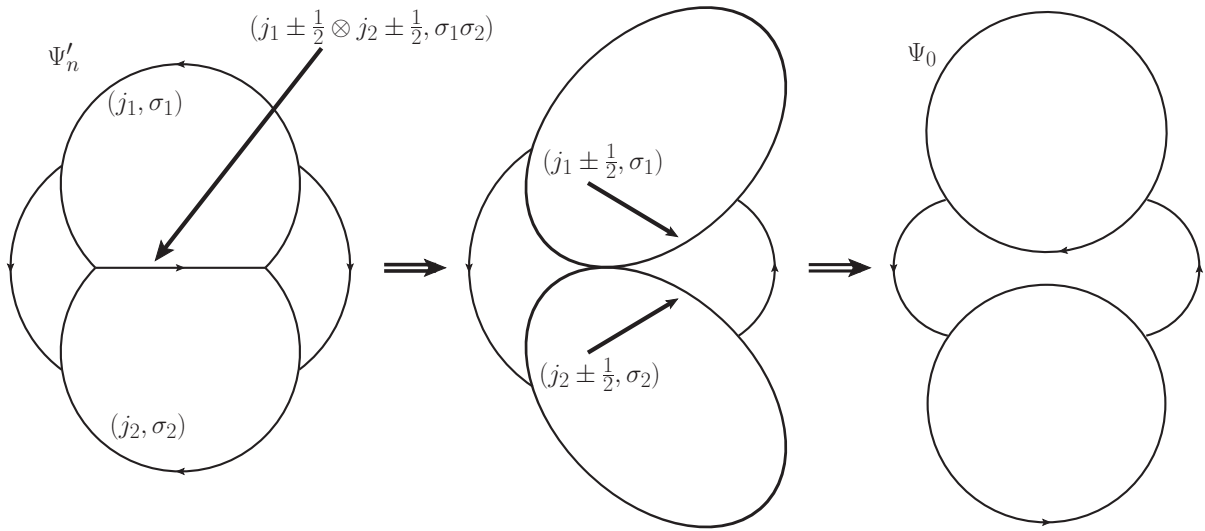}
\caption{By making the move V' twice on this topspin network we can remove two extraordinary edges. Again, we are suppressing intertwiners.}
\label{fig:example}
\end{center}\end{figure}

Now consider the effect of this move on a specific topspin network, figure \ref{fig:example}. The topspin-net $\Psi_n'$ has two extraordinary edges on the sides and the center line has the specific representation denoted in the diagram. We then use the move $V'$ on the right typical vertex, making the two extraordinary edges share a typical vertex. Now perform the move $V_2$ on this typical vertex, removing it. The definition of extraordinary edges requires that edges that come from the extraordinary vertices of the extraordinary edges must intersect at least once, at most twice. But in figure \ref{fig:example} they \textit{never} intersect, and so we have removed the extraordinary edges.

So we have explicitly shown that at least in some cases, there can be two different topspin networks which are equivalent under geometric covering moves but which are not of the same level. In fact, it would also appear they are from different source networks as well. In figure \ref{fig:example}, $\Psi_0$ is a source topspin-net (no extraordinary edges) but $\Psi_n'$ appears to have a source net like the $\theta$-graph. In other words, we have shown theorem \ref{thm:disjoint} does not hold for topspin-nets.

The statement is now that the categories $\mathcal{C}_{\mathcal{H}}(\psi,G),~\mathcal{C}_{\mathcal{H}}(\Psi,\mathcal{G})$ cannot be the same since it is clearly not possible to find a structure-preserving functor between them. Even at a very abstract level, the structure of topspin networks under the action of the Hamiltonian is quite different than in the usual case of spin networks. The most obvious consequence that follows from this observation is that the kernel of the Wheeler-DeWitt operator no longer reduces to a linear problem \cite{TTbook}. 

We should also note that the category $\mathcal{C}_{\mathcal{H}}(\Psi,\mathcal{G})$ is \textit{not the same} as the loop quantum gravity 2-category $\mathcal{L}(G)$ constructed in \cite{DMA}. However, the 1-morphisms in that category are in fact the equivalence relations we imposed on our category. In principle, one could check to see if these two categories are compatible with each other and enlarge $\mathcal{L}(G)$ to include the action of the Hamiltonian. However, this construction would play no part in our current analysis so we leave it to a later work.

\section{Conclusion}\label{s:conclusion}
In this paper we have expanded on the proposal that topological information can be included in the framework of loop quantum gravity by using topspin networks. This idea is based on the fact the spatial sections of a foliated spacetime can be realized as a branched covering space. If we identify the branch locus with a spin network and add topological labels, we should be able to study both the topology and the geometry which results from the canonical quantization of gravity.

We implemented this by requiring that the metric be invariant under a deck transformation over the branch locus. This extends the structure group of gravity to a $SU(2)\times S_n$ gauge theory. This symmetry naturally applies to the phase space as well, the only change being that the fields now take values in the universal enveloping algebra $\mathcal{U}(\mathfrak{su}(2))$, rather than $\mathfrak{su}(2)$. This is naturally seen as a generalization of the usual LQG case, since there is an injection $i:\mathfrak{su}(2)\hookrightarrow \mathcal{U}(\mathfrak{su}(2))$. It can also be viewed as promoting the fields to operators from the outset and allowing them to take polynomial values.

We found that this change does not greatly impact the structure of the theory by explicitly showing the canonical transformations of the phase space and the Hilbert space of cylindrical functions retain most of their structure. Thus, all of the techniques known to the LQG community about quantizing background-independent theories should be applicable in this case as well. The inner product is not gauge invariant on the entire Hilbert space $\mathcal{K}$, but it is on a subset $\mathcal{K}_0$ containing the gauge invariant spin networks.

We took the familiar case of the area operator and showed that in our construction, the usual result must be multiplied by the order of the symmetry group over the branch locus. One could write this as a set of integers $(m_1,...,m_L)$ for $L$ arcs of the graph, where $m_k!=|S_{m_k}|$. This does not alter the qualitative features of the theory; the area operator still measures the geometric area by counting the spin labels. 

We also discussed the action of the Hamiltonian on topspin networks, and how it would be different on cylindrical functions of these networks. Using the category-theoretic viewpoint, it is clear that the two situations are inequivalent, and the inclusion of topological information complicates the situation. This is naturally due to the fact that there may be more than one topologically equivalent way to specify the spatial sections as branched covering spaces. In the same way that consideration of gauge symmetry complicates quantum field theory, this topological symmetry naturally complicates this analysis. At this point we do not discuss what would have to be done to solve the Hamiltonian constraint in this approach, but hopefully we have displayed the key differences between using spin networks and topspin networks in LQG.

\subsection*{Topology Change in LQG}
We close with a few words on the status of topology change in LQG. Now that we have determined how the area operator and the Hamiltonian operator act on topspin networks, we can try to answer the question, ``does topology change occur in LQG?'' At the level of this analysis, the answer is yes.

It is clear from \S\ref{s:Area_Operator} that the area operator is capable of tracking some changes in the topology, because it explicitly contains information regarding the branch locus. This is not a complete specification, since there is apparently nothing to prevent other branch loci from having the same identifications $m_I$ but different topologies - identifying different sheets or even having a different covering degree. In fact, the area operator is of somewhat limited use when studying topology change, since the area of a surface intersecting the spin network might not be preserved under a change of topology anyway. However, it does illustrate the relationship between topology and geometry; the area of a surface can depend on the topology if you change the topology in such a way that the surface changes. This would occur, for instance, in the well-known trouser topology \cite{Anderson-DeWitt}. As the spatial section passed from the trunk of the pants to the legs, if the surface $S$ was punctured by the critical point in the crotch, the area of the surface near the critical point could clearly change.

On the other hand, the Hamiltonian generally appears to directly change the spatial topology by adding edges which might not be removable with geometric covering moves. Of course, we have not included the complete dynamics in our analysis, so this statement is only true at the level of kinematics. The question also appears to be difficult to answer completely since it would require a complete characterization of the topspin networks from the perspective of geometric covering moves. Showing two topspin networks are equivalent reduces to just finding one set of covering moves; showing they are inequivalent is showing there is \textit{none}. Such partial answers are a common feature in topology, but there is certainly nothing to suggest one will \textit{always} be able to find a covering move to undo the action of the Hamiltonian. Thus, at the level of kinematics this approach admits topology change.

This is clearly not the only proposal in the literature which attempts to track topological information in quantum gravity, but we believe it represents an interesting one. The construction naturally follows from Alexander's theorem and nicely fits into the existing theory of loop quantum gravity. The connection to the topology through the fields and operators allows one to formulate questions regarding it, which is not generally true in LQG since the reduction of the 3-manifold to a graph destroys the topological structure. If we had the full dynamical theory of LQG one could begin to carefully study the inclusion of topology by using the methods discussed here. This could lead to a theory of quantum gravity which is topologically relative as well as geometrically relative, and could give us some insight into the topology of our spacetime models.

\ack
I would like to acknowledge M. Marcolli for her great assistance with this project. Without her this work would not have been possible. The comments from the two reviewers were also very valuable in clarifying several important points.

\appendix

\section{Example of Different Topologies Over a Common Branch Locus}\label{ap:knots}
This example comes from the book \cite{PS}. Specialize to \textit{cyclic} branched coverings $\pi:\mathbb{S}^3\to\mathbb{S}^3$. Thinking of the 3-sphere as $\mathbb{R}^3\cup\infty$, the cyclic action is obtained by identifying points of a rotation by $2\pi/n$ about a line $l\subset\mathbb{R}^3$. The rotation group here is $\mathbb{Z}_n$, and by performing surgery on a curve in the base space and the preimage of that curve in the branched surface we can construct two inequivalent topologies branched over a common locus.

First we recall the definition of a framed knot. 

\begin{definition}
A \textbf{framed knot} is an embedding $K:\mathbb{S}^1\hookrightarrow\mathbb{S}^3$ together with a specification of the trivialization of the normal bundle $N_{\mathbb{S}^3\setminus K}$. This framing can be completely described with an integer which tells us how many times (and in what direction) the normal bundle wraps around the knot.
\end{definition} 

\begin{figure}
\begin{minipage}[b]{0.5\linewidth}
\begin{center}\leavevmode
\xygraph{
	!{0;/r2.0pc/:}
	!{\hover}
	!{\hcap}
	[l]!{\vcap[3]}
	[rrr]!{\xcapv@(0)=<}
	[lll]!{\vcap[-3]}
}
\caption{A framed knot describes a ribbon with a twist.}\label{fig:twist}
\end{center}
\end{minipage}
\begin{minipage}[b]{0.5\linewidth}
\begin{center}\leavevmode
\xygraph{
	!{0;/r4.0pc/:}
	!{\xunderv[-1]=>>{-1}}
	[rru]!{\xoverv[-1]=>>{+1}}
}
\caption{When a knot crosses itself the framing is changed by one. This can wrap (+1) or unwrap (-1) the framing.}\label{fig:crossing}
\end{center}
\end{minipage}

\end{figure}

It turns out that using the belt trick the framing of a knot can be described by twists given by the basic Reidemeister move. Figure \ref{fig:twist} shows $\mathbb{S}^1$ with a $+1$ framing, and figure \ref{fig:crossing} shows the sign convention for the framing of the normal bundle. Specifically, take the knot $J$ in figure \ref{fig:base}, which has framing +1: two positive crossings on the right and one negative crossing on the left. In each of the cyclic covers, the inverse image of this knot has $n$ copies of the double crossing (see figure \ref{fig:cover} for $n=2$ case). It is pretty easy to see that in the $n$-fold cover the preimage will be $n$ curves with framing -1 linked as shown.

\begin{figure}
\begin{minipage}[b]{0.5\linewidth}
\begin{center}\leavevmode
\xygraph{
	!{0;/r1.5pc/:}
	!{\hover}
	!{\hover-}
	[llllll]!{\vcap[6]}
	!{\hunder}
	!{\hcap}	
	[ld]!{\vcap[-6]=<}
	[rr]!{\vcap[-2]}
	[u]!{\xcapv@(0)}
	[u]!{\vcap[2]=>}
}
\caption{A knot $J$ with framing $+1$ in the base space.}
\label{fig:base}
\end{center}
\end{minipage}
\begin{minipage}[b]{0.5\linewidth}
\begin{center}\leavevmode
\xygraph{
	!{0;/r4.0pc/:}
	!P8"a"{~:{(0.6,0):}~>{}}!P16"c"{~:{(1.4,0):}~={78.75}~>{}}!P20"g"{~:{(1.1,0):}~={9}~>{}}!P16"e"{~:{(1.4,0):}~>{}}!P20"f"{~:{(1.8,0):}~={17}~>{}}!P12"b"{~>{}}!P20"h"{~:{(0.85,0):}~={9}~>{}}
	!{\hover~{"a1"}{"b1"}{"a8"}{"b12"}}
	!{\hunder~{"c14"}{"b1"}{"c13"}{"b12"}}
	!{\vcap~{"f2"}{"f3"}{"c14"}{"c1"}=>}
	!{\vover~{"c1"}{"c2"}{"g5"}{"g6"}}
	!{\vcap~{"f7"}{"f8"}{"c2"}{"c5"}}
	!{\hover~{"c5"}{"b6"}{"c6"}{"b7"}}
	!{\hunder~{"a4"}{"b6"}{"a5"}{"b7"}}
	!{\vcap~{"f12"}{"f13"}{"c6"}{"c9"}=>}
	!{\vover~{"c9"}{"c10"}{"g15"}{"g16"}}
	!{\vcap~{"f17"}{"f18"}{"c10"}{"c13"}}
	!{\vcap~{"e6"}{"e4"}{"a4"}{"a1"}=>}	
	!{\vcap~{"e12"}{"e14"}{"a5"}{"a8"}=<}
	!{\vcap~{"h5"}{"h6"}{"g5"}{"g6"}}
	!{\vcap~{"h15"}{"h16"}{"g15"}{"g16"}}
}

\caption{The preimage of the curve in a 2-fold branched cover, a pair of linked knots $J_1$ and $J_2$ with framing $-1$ each. Extending to $n$ covers results in $n$ links, each with framing $-1$.}
\label{fig:cover}
\end{center}
\end{minipage}
\end{figure}

This framing is all that is required to define \textit{rational surgery} (or Dehn surgery) of a 3-sphere, in which a torus is removed and then reattached by twisting a number of times around the meridian given by the framing. By the Hilden-Montesinos theorem, any closed oriented 3-manifold can be obtained in this way \cite{PS,GS}. 

The linking $\{J,K\}$ with $K=l\cup\infty$ actually describes the Whitehead link, which is a trefoil ($K$) linked with the unknot ($K$) with framing $+1$. We now perform this surgery in the base space by taking an $\epsilon$-neighborhood about the knot $J$ from figure \ref{fig:base}. Since the 3-sphere can be described by the gluing of two solid tori, where the parallel of one is the meridian of the other, this surgery returns our base space to $\mathbb{S}^3$.

Next, it is pretty easy to see that the link $\{J_1,J_2\}$ in $n=2$ (figure \ref{fig:cover}) is isomorphic (through Reidemeister moves) to $\mathbb{S}^1$ with framing +3. This is a generalization of the above process, and it creates the lens space $L(3,1)=\mathbb{S}^3/ \mathbb{Z}_3$ (note $\mathbb{S}^3=L(1,0)$).

By proposition 22.9 of \cite{PS}, the $n=5$ case is actually a Poincar\'e sphere $\mathbb{S}_H^3$. These two spaces are not homeomorphic, although they are both branched over the trefoil in $\mathbb{S}^3$. This can be seen from their homology groups. The Poincar\'e sphere has the homology of a sphere,
\begin{equation}H_k(\mathbb{S}_H^3)=\Biggl\lbrace \begin{array}{ccc}
\mathbb{Z}&\quad&k=0,3\\
0&&k=1,2.\end{array}\end{equation}
However, the Lens space has a nontrivial middle homology group which arises from the action of $\mathbb{Z}_3$ on the unknot:
\begin{equation}H_k(L(3,1))=\Biggl\lbrace\begin{array}{ccc}
\mathbb{Z}&\quad&k=0,3\\
\mathbb{Z}_3&&k=1\\
0&&k=2\end{array}\end{equation}
Thus we have two inequivalent topological spaces which have the same branch locus in $\mathbb{S}^3$.

\section{An Example of the Holonomy and Cylindrical Functions}\label{ap:ex}
\begin{figure}\begin{center}
\includegraphics[scale=0.4]{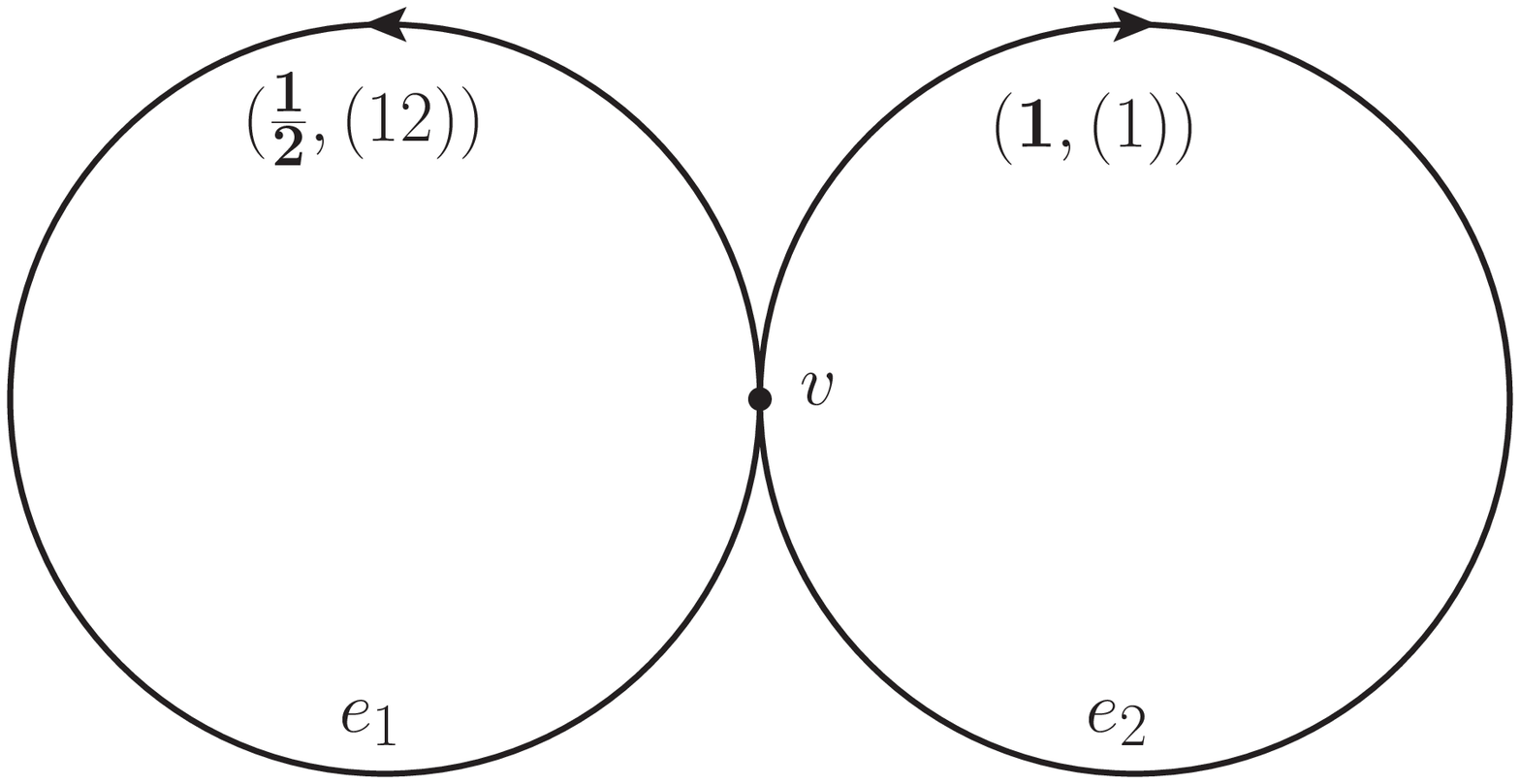}
\caption{The eyeglasses graph as an example topspin network.}\label{fig:Eyeglass}
\end{center}\end{figure}

To illustrate some of the ideas presented in this paper we will give an example which shows the form of the holonomy and the cylindrical functions on a simple topspin network. Consider the ``eyeglasses'' graph (figure \ref{fig:Eyeglass}), which consists of two arcs $e_1$, $e_2$ and one vertex $v$ with appropriate coloring for a topspin network. We assume this is the branch locus of an order 3 covering.

We take $T_i$ as generators of $\mathfrak{su}(2)$, with a dimension $\lambda=2j+1$ representation $d^{\lambda}(T_i)=d^{(\lambda)}_i$, and $\sigma$ as elements of $S_n$. The connection is a function of the elements of $S_3$, taking values in $\mathcal{U}(\mathfrak{su}(2))$-valued one-forms,
\begin{equation}
A(\sigma)=A_a^j(\sigma)\tau_jdx^a.
\end{equation}
There is also a dimension $\lambda+1$ representation of the universal enveloping algebra, $\pi_{\lambda}(\tau_i)=\pi_i^{(\lambda)}$. Over the $I$th edge the holonomy as a function of $\sigma\in S_3$ is
\begin{eqnarray}
h(A(\sigma),e_I(s))&=\exp_{\lambda_I}\left(\int_0^1A^j_a(\sigma)\tau_j\dot{e}^a_Ids\right)\\
&=\exp_{\lambda_I}\left(\mathcal{I}^j(A(\sigma),e_I)\tau_j\right).
\end{eqnarray}
Here $\dot{e}^a=dx^a/ds$ and we have the constant (with respect to $\mathcal{U}(\mathfrak{su}(2))$) integral $\mathcal{I}^j(A(\sigma),e_I)=\int_0^1A(\sigma)^j_a\dot{e}^ads$. Now using the definition of our exponential map (\ref{eq:exp_l}) we have
\begin{eqnarray}
\exp_{\lambda_I}\left(\mathcal{I}^j(A(\sigma),e_I)\tau_j\right)&=\exp\circ\theta_{\lambda_I}(\mathcal{I}^j(A(\sigma),e_I)\tau_j)\\
&=\exp\left(\mathcal{I}^j(A(\sigma),e_I)\theta_{\lambda_I}(\tau_j)\right),
\end{eqnarray}
where $\theta_{\lambda_I}(\tau_j)$ would be the element in the matrix Lie algebra $\mathfrak{gl}_{\lambda_I}(\mathbb{R})$ representing the appropriate polynomial $\tau_j$ of basis elements of $\mathfrak{su}(2)$. For example, in the spin-1/2 representation with generators $T_j=i\sigma_j$ ($\sigma_3$ diagonal), the element $T_1T_2T_3^2$ would be represented by
\begin{eqnarray}
\pi_1(T_1T_2T_3^2)&=\pi_1(\rmi\sigma_1)\pi_1(\rmi\sigma_2)[\pi_1(\rmi\sigma_3)]^2\\
&=\rmi\left(\begin{array}{cc}
1&0\\
0&-1\end{array}\right)=i\pi_1(\sigma_3).
\end{eqnarray}
Over the first edge in figure \ref{fig:Eyeglass} the holonomy would be
\begin{equation}
\fl H(A,e_1(s))=\exp\left(\mathcal{I}^j(A(\sigma_1,e_1))\theta_1(\tau_j)\right)\times...\times\exp\left(\mathcal{I}^j(A(\sigma_6,e_1))\theta_1(\tau_j)\right).
\end{equation}
Now enforcing the symmetry given by the permutation label $(12)$ on $e_1$, we reduce the group from $S_3\rightarrow S_2$ and find
\begin{equation}
H(A,e_1(s))=\exp\left(\mathcal{I}^j(A(\sigma_1),e_1)\theta_2(\tau_j)\right)\times\exp\left(\mathcal{I}^j(A(\sigma_2),e_1)\theta_2(\tau_j)\right)
\end{equation}
where $H(A,e_1) \in \mathcal{G}_1=GL_{2}(\mathbb{R})\times GL_2(\mathbb{R})$. The edge $e_2$ has trivial symmetry label, so the holonomy would be
\begin{equation}
\fl H(A,e_2(s))=\exp\left(\mathcal{I}^j(A(\sigma_1),e_2)\theta_2(\tau_j)\right)\times...\times\exp\left(\mathcal{I}^j(A(\sigma_6),e_2)\theta_2(\tau_j)\right)
\end{equation}
and $H(A,e_2) \in \mathcal{G}_2=GL_3(\mathbb{R})^6$. The entire group is $\mathcal{G}=\mathcal{G}_1\times \mathcal{G}_2$.
The groupoid $l$ for this graph contains one object, the vertex $v$ and two morphisms $\alpha$, $\beta$ with source and target maps
\begin{equation}
t(\alpha)=s(\alpha)=v,~t(\beta)=s(\beta)=v.
\end{equation}
This groupoid does not have any non-trivial subgroupoids. The cylindrical functions are given by maps on the elements of $\mathcal{G}$:
\begin{equation}
f(x)=\Psi(H(A,e_1)\times H(A,e_2)),~x:l\to \mathcal{G}.
\end{equation}

\section*{References}


\end{document}